\documentclass[article, twocolumn,
   aps, pra,
  amsmath,amssymb,
  longbibliography,
  ]{revtex4-2}
\usepackage{graphicx,color}
\usepackage{amsmath}
\usepackage{natbib}
\usepackage{epsfig}
\usepackage{comment}
\begin{document}

\title{\textcolor{blue}{Variational approach to Yukawa fluids. II. Instantaneous elastic moduli and sound velocities.}}

\author{S. A. Khrapak}\email{Sergey.Khrapak@gmx.de} 
\author{A. G. Khrapak}

\affiliation{Joint Institute for High Temperatures, Russian Academy of Sciences, 125412 Moscow, Russia}

\begin{abstract}
 The variational approach based on the Bogoliubov inequality using the fluid of hard spheres as a reference system is implemented to evaluate instantaneous shear, bulk and longitudinal elastic moduli, as well as related sound velocities of Yukawa fluids. The remarkable accuracy of this method is documented. In addition, we evaluate the adiabatic sound velocity from an appropriate equation of state and discuss its relation to the longitudinal and bulk sound velocities obtained from the corresponding instantaneous elastic moduli. The transition between weakly coupled and strongly coupled regimes is analyzed in detail. 
\end{abstract}

\date{\today}

\maketitle

\section{Introduction}

It is well established that when a perturbation is applied suddenly to a fluid, it responds elastically at first, just like a solid~\cite{ZwanzigJCP1965}. This initial response is conventionally described by high-frequency (instantaneous) elastic moduli.
Elastic moduli play an extremely important role in the physics of fluids, affecting and regulating many properties related to their dynamics and thermodynamics. For example, elastic moduli determine the longitudinal and transverse velocities of sound-like collective excitations in dense fluids at high frequencies and short wavelengths (elastic regime)~\cite{Hubbard1969,BalucaniBook,MorkelPRE1993,GoldenPoP2000,KawPoP2001,AliottaPRE2011,BrykPRE2014,HosokawaJPCM2015,KhrapakSciRep2017,KhrapakPRE05_2021}; the Maxwellian relaxation time is inversely proportional to the instantaneous shear modulus~\cite{MountainJCP1966,TrachenkoBook,HartkampPRE2013,KhrapakPRE11_2024}; instantaneous elastic moduli regulate the coefficients of self-diffusion and thermal conductivity within the vibrational paradigm of transport properties of dense fluids~\cite{KhrapakPRE01_2021,KhrapakPoP08_2021,KhrapakMolecules12_2021,KhrapakPRE12_2023,KhrapakPhysRep2024}, as well as the value of the  Stokes-Einstein product in the relation between the self-diffusion and shear viscosity coefficients~\cite{ZwanzigJCP1983,KhrapakMolPhys2019,KhrapakPRE10_2021}; the excess entropy of dense fluids with repulsive interactions can be relatively accurately estimated from the known elastic moduli~\cite{KhrapakJCP2021,KhrapakPRE09_2024}; the magnitude of the shear elastic modulus can serve as an indicator of the gas-like to fluid-like (Frenkel) crossover~\cite{HuangPRR2023,YuPRE2024,XuPRR2026,KhrapakPoF2026}.

The high-frequency elastic moduli of simple (monatomic) fluids can be expressed in terms of the
pair-interaction potential $\phi(r)$ and the radial distribution function (RDF) $g(r)$. A detailed derivation was presented by Zwanzig and Mountain~\cite{ZwanzigJCP1965}. The expressions obtained for the instantaneous shear modulus ($G_{\infty}$) and the bulk modulus ($K_{\infty}$) are 
\begin{equation}\label{G_}
G_{\infty}=nT+\frac{2\pi n^2}{15}\int_0^{\infty}dr r^3 g(r)\left[r \phi''(r)+4\phi'(r)\right],
\end{equation}
\begin{equation}\label{K_}
K_{\infty}=\frac{5}{3}nT+\frac{2\pi n^2}{9}\int_0^{\infty}dr r^3g(r)\left[r \phi''(r)-2\phi'(r)\right].
\end{equation}
Here $n$ is the number density, $T$ is the temperature expressed in energy units ($=k_{\rm B}T$), and $\phi'(r)$ and $\phi''(r)$ are the first and second derivatives of the interaction potential with respect to distance $r$. The first terms in each of the equations above correspond to the kinetic (ideal gas) contribution, the second terms are the excess (configurational) parts. To evaluate the configurational contribution, knowledge of the RDFs is generally required. The latter can be obtained from Monte Carlo (MC) or molecular dynamics (MD) numerical simulations~\cite{HeyesJCP1994,BamdadChemPhys2006,HeyesJCP2008,KhrapakSciRep2017,KhrapakPoP02_2020}
or from the integral equation of state theory (Ornstein–Zernike equation supplemented by an appropriate closure)~\cite{RogersPRA1984,KalmanPRL2000,ToliasPoP2021}. However, in some cases, explicit knowledge of RDFs is not required. For example, for the trivial case of inverse-power-law interactions, the elastic moduli can be expressed using the excess energy or pressure of the system~\cite{KhrapakSciRep2017}. In the case of conventional Lennard-Jones ($12-6$) and generalized ($n-m$) Lennard-Jones (Mie) potentials, elastic moduli can be related to a combination of excess pressure and internal energy of the fluid~\cite{ZwanzigJCP1965,Meier2002,KhrapakMolecules2020,KhrapakMolecules2021}. Moreover, for soft repulsive interactions such as Coulomb and screened Coulomb potentials, simplistic models of RDF, such as step-wise and two-step approximations, have proven to be surprisingly accurate~\cite{GoldenPSS1993,KhrapakPoP2016,KhrapakAIPAdv2017,KhrapakIEEE2018,FairushinResPhys2020,FairushinFluids2023}.  

The purpose of this paper is to examine another situation in which explicit knowledge of RDF is not required. We focus on the repulsive screened Coulomb interaction potential (also known as Debye-H\"uckel and Yukawa potential), which is widely used in the context of plasma-related systems (dusty and complex plasmas) and colloidal suspensions~\cite{TsytovichUFN1997,ShuklaRMP2009,FortovPR,MorfillRMP2009,BonitzRPP2010,FortovBook,IvlevBook,MelzerBook}. A variational approach based on the Bogoliubov inequality with the hard-sphere (HS) model chosen as the reference system has a long history of application to estimate the thermodynamic properties of Coulomb and Yukawa strongly coupled fluids. In a companion paper we have demonstrated that Percus-Yevick (PY) RDF combined with virial expression for excess entropy of the HS fluid (as previously suggested by deWitt and Rosenfeld for one-component plasma~\cite{DeWittPLA1979}) are particularly appropriate for the  variational calculation of thermodynamic properties of Yukawa fluids and quantified the (high) precision of this method~\cite{YukawaSubmitted}. Here we go further, by demonstrating that a variational approach is applicable to evaluate the instantaneous elastic moduli as well. Using the variational procedure and the known analytical properties of the PY RDF, the integrals in Eqs.~(\ref{G_}) and (\ref{K_}) can be expressed in a simple analytical form. The accuracy of this procedure is verified. We use this opportunity to illustrate the relations between the sound velocities defined by the instantaneous elastic moduli and the adiabatic sound velocity of the Yukawa fluid. The relations with the celebrated concept of dust acoustic velocity in dusty plasmas are discussed.                

\section{Main definitions}

We consider a one-component system of particles interacting via the pairwise Yukawa repulsive potential of the form
\begin{equation}\label{Yukawa}
\frac{\phi(r)}{T}=\frac{\Gamma a}{r}\exp(-\kappa r/a),
\end{equation}
where $\Gamma=Q^2/aT$ is the coupling parameter,  $\kappa = a/\lambda$ is the screening parameter, $Q$ is the particle charge,  $a=(4\pi n/3)^{-1/3}$ is the Wigner-Seitz radius, $n$ is the number density, $\lambda$ is the screening length, and $T$ is the temperature expressed in energy units ($\equiv k_{\rm B}T$).
Further details on the properties of the Yukawa fluid can be found in the companion paper~\cite{YukawaSubmitted}.

Let us introduce the dimensionless thermodynamic functions of Helmholtz free energy, energy, entropy and pressure: $f=F/NT$, $u=U/NT$, $s=S/Nk_{\rm B }$, and $p=P/nT$, where $N$ is the number of particles. They can be presented as sums of the ideal gas (subscript ``id'') and excess (subscript ``ex'') contributions: $f=f_{\rm id}+f_{\rm ex}$, $u=3/2+u_{\rm ex}$, $s=s_{\rm id}+s_{\rm ex}$, $p=1+p_{\rm ex}$. The excess contributions are related to particle-particle interactions. Expressed in terms of excess contributions, the Bogoliubov inequality with HS fluid chosen as a reference system can be formulated as~\cite{YukawaSubmitted}
\begin{equation}\label{Bogoliubov2}
 f_{\rm ex}\leq -s_{\rm ex}(\eta)+u_{\rm ex}(T,\eta). 
\end{equation}
where $s_{\rm ex}(\eta)$ is the entropy of the HS fluid and $u_{\rm ex}(T,\eta)$ is the excess energy of the Yukawa fluid evaluated using the appropriate RDF of the HS fluid. Both quantities depend on the conventional packing fraction of the HS fluid $\eta$, which is considered a variational parameter. Minimizing the right-hand side (RHS) of Eq.~(\ref{Bogoliubov2}) with respect to $\eta$ yields the best estimate of the free energy of the Yukawa fluid. This is the essence of the method, and in the companion paper we applied it using the PY RDF of the HS fluid along with the virial expression for the excess entropy. We have demonstrated the high accuracy of this method with respect to calculation of the excess energy, entropy and pressure~\cite{YukawaSubmitted}. The idea here is to use this method to evaluate other related integrals involved in the expressions for instantaneous elastic moduli of Yukawa fluids.

\section{Elastic moduli and sound velocities}

\begin{widetext}
Using the PY RDF $g_0(r;\eta)$ and expressing the integrals in Eqs.~(\ref{G_}) and (\ref{K_}) in dimensionless units, the reduced elastic moduli become
\begin{equation}\label{shearmod}
\frac{G_{\infty}}{nT}= 1-\frac45\Gamma\eta^{2/3}\int_0^{\infty}xe^{-tx}(1+tx-\frac12t^2x^2)g_0(x;\eta)dx= 1-\frac45\Gamma\eta^{2/3}\left[G(t,\eta)-t\frac{\partial G(t,\eta)}{\partial t}-\frac{t^2}{2}\frac{\partial^2  G(t,\eta)}{\partial t^2}\right]   
\end{equation}
and 
\begin{equation}\label{bulkmod}
\frac{K_{\infty}}{nT}= \frac53+\frac83\Gamma\eta^{2/3}\int_0^{\infty}xe^{-tx}(1+tx+\frac14t^2x^2)g_0(x;\eta)dx=  \frac53+\frac83\Gamma\eta^{2/3}\left[G(t,\eta)-t\frac{\partial G(t,\eta)}{\partial t}+\frac{t^2}{4}\frac{\partial^2  G(t,\eta)}{\partial t^2}\right]. 
\end{equation}
Here $x=r/\sigma$, $\sigma$ is the diameter of hard spheres and $t=\sigma/\lambda=2\eta^{1/3}\kappa$. The function $G(t,\eta)$ is known analytically~\cite{Wertheim1963,Thiele1963}
\begin{equation}
G(t,\eta)=\frac{tL(t,\eta)}{12\eta[L(t,\eta)+S(t,\eta)e^t]},
\end{equation}
where 
\begin{equation}
L(t,\eta)= 12\eta[(1+\tfrac{1}{2}\eta)t+(1+2\eta)]
\end{equation}
and 
\begin{equation}
S(t,\eta)=(1-\eta)^2t^3+6\eta(1-\eta)t^2+18\eta^2 t-12\eta(1+2\eta).
\end{equation}

Furthermore, the shear and bulk moduli can be combined into the longitudinal modulus
\begin{equation}\label{longmod}
\frac{M_{\infty}}{nT}=\frac{K_{\infty}}{nT} +\frac43 \frac{G_{\infty}}{nT} =3+\frac85\Gamma\eta^{2/3}\left[G(t,\eta)-t\frac{\partial G(t,\eta)}{\partial t}+\frac{3t^2}{4}\frac{\partial^2  G(t,\eta)}{\partial t^2}\right]. 
\end{equation}
\end{widetext}
Thus, the elastic moduli can be expressed from the fully analytical function $G(t,\eta)$ and its first two derivatives with respect to $t$. The relation between $\Gamma$ and $\eta$ is provided by the variational procedure as explained in detail in the companion paper~\cite{YukawaSubmitted}.

Note that by appropriately combining each pair of elastic moduli one can get rid of the second derivative of $G(t,\eta)$ and obtain the relation to the excess pressure.  
\begin{equation}\label{CaushiRel}
\frac{M_{\infty}}{nT}- 3 \frac{G_{\infty}}{nT} =2p_{\rm ex}. 
\end{equation}
This is known as a generalized Cauchy identity in the theory of elasticity~\cite{ZwanzigJCP1965}.

We can further define the longitudinal, transverse, and bulk sound velocities through the corresponding elastic moduli~\cite{BalucaniBook,KhrapakSciRep2017,KhrapakPoF2023}:
\begin{equation}
M_{\infty}=mnc_l^2, \quad G_{\infty}=mnc_t^2, \quad K_{\infty}=mnc_b^2,    
\end{equation}
where $m$ is the mass of the particles. We have calculated the longitudinal and transverse sound velocities $c_l$ and $c_t$ to compare with the results existing in the literature. The results are summarized in Tab.~\ref{Tab2}. The first two columns specify the location of the system state point in the $(\kappa,\Gamma)$ plane. The next two columns list the longitudinal and transverse sound velocities evaluated within the quasi-localized charge approximation (QLCA) framework~\cite{GoldenPoP2000,KalmanPRL2000,DonkoJPCM2008} using RDFs from direct MD simulations in Ref.~\cite{KhrapakPoP2016}. The results were tabulated in Tab. II of Ref.~\cite{KhrapakPoP02_2016_1}. The next two columns correspond to the longitudinal and transverse velocities calculated here using the variational approach. Only the configurational contribution is considered, i.e., kinetic terms in elastic moduli are omitted, as usually done within the QLCA approach~\cite{GoldenPoP2000,KalmanPRL2000,DonkoJPCM2008}. This is convenient at sufficiently strong coupling, when the coupling parameter exceeds that of the  gas-like to liquid-like dynamical crossover, referred to as the Frenkel line on the phase diagram~\cite{BrazhkinPRE2012,BrazhkinUFN2012,BrazhkinPRL2013}. For Yukawa fluids with $\kappa\lesssim 3$ this crossover occurs around $\Gamma/\Gamma_{\rm m}\simeq 0.05$~\cite{HuangPRR2023,YuPRE2024}. All velocities in Tab.~\ref{Tab2} are expressed in units of thermal velocity $v_{\rm T}=\sqrt{T/m}$.  The table demonstrates excellent agreement between the elastic sound velocities obtained from the QLCA approach with RDFs from a direct MD simulation and the variational method implemented here. 

\begin{table}
\caption{\label{Tab2} Configurational components of the longitudinal and transverse sound velocities tabulated in Tab.~II of Ref.~\cite{KhrapakPoP02_2016_1} (superscript ``QLCA'') and calculated here using the variational approach (superscript ``Var''). The first two columns specify the location of the system state point on the $(\kappa,\Gamma)$ plane. The velocities are expressed in units of the thermal velocity $v_{\rm T}$. }
\begin{ruledtabular}
\begin{tabular}{llrrrr}
$\kappa$ & $\Gamma$ & $c_l^{\rm QLCA}$ &  $c_t^{\rm QLCA}$ & $c_l^{\rm Var}$ & $c_t^{\rm Var}$   \\ \hline
0.5 & 145 & 41.29 & 3.98 & 41.30 &  4.00  \\
1.0 & 180 & 22.29 & 4.07 & 22.28 & 4.08   \\
2.0 & 370 & 13.83 & 4.26 & 13.78 & 4.26 \\
3.0 & 990 & 11.55 & 4.41 & 11.53 & 4.42  \\
\end{tabular}
\end{ruledtabular}
\end{table}

\section{Strong coupling limit}

In the strong coupling limit, the elastic moduli and elastic sound velocities are dominated by the configurational contribution, the kinetic terms being negligibly small. The configurational contributions are very nearly proportional to the coupling parameter $\Gamma$. This implies that in properly reduced units the elastic moduli and related sound velocities become functions of the screening parameter $\kappa$ alone~\cite{KalmanPRL2000,KhrapakPoP10_2019}. We can derive remarkably simple analytical expressions appropriate in this limit using variational consideration. 

As we have discussed in the companion paper~\cite{YukawaSubmitted}, the excess energy from the PY theory evaluated at an unphysical packing fraction $\eta=1$ has a very special meaning. It defines the fluid Madelung energy~\cite{RosenfeldPRE2000},
represents a rather accurate approximation of the excess energy. This gives hope that other quantities expressed with the help of integrals involving $g_0(x,\eta)$ can be relatively accurately estimated using $g_0(x;1)$ in the strongly coupled regime.

We have applied this procedure to calculate the instantaneous elastic moduli and the sound velocities of the Yukawa fluid. This is relatively straightforward and requires manipulations with the function $G(t,\eta)$ and its derivatives. The results expressed in terms of the reduced instantaneous sound velocities is  
\begin{equation}\label{elasticsound_general}
c_{l.t}=\frac{\omega_p a}{3\sqrt{5}}{\mathcal{F}_{l,t}(\kappa)},
\end{equation}
where $\omega_p=\sqrt{4\pi n Q^2/m}$ is the plasma frequency and the $\kappa$-dependent functions are 
\begin{equation}\label{longitudinal}
\mathcal{F}_{l}^2(\kappa) = \frac{\kappa^4\left[(4+3\kappa^2)\sinh(\kappa)-4\kappa\cosh(\kappa)\right]}{\left[\kappa\cosh(\kappa)-\sinh(\kappa)\right]^3}    
\end{equation}
and
\begin{equation}\label{transverse}
\mathcal{F}_{t}^2(\kappa) = \frac{\kappa^4\left[(3+\kappa^2)\sinh(\kappa)-3\kappa\cosh(\kappa)\right]}{\left[\kappa\cosh(\kappa)-\sinh(\kappa)\right]^3}. 
\end{equation}
The small kinetic terms have been omitted at strong coupling, as discussed above. Similar expressions have previously been obtained in Ref.~\cite{KhrapakPoP10_2019} using a different line of arguments. The sound velocities can be easily expressed in units of the thermal velocity using the identity $\omega_{\rm p}a/v_{\rm T}=\sqrt{3\Gamma}$.

\begin{figure}
\includegraphics[width=7cm]{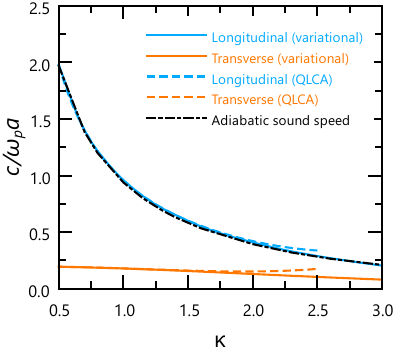}
\caption{(Color online) The reduced instantaneous longitudinal and transverse elastic sound velocities in the strongly coupled Yukawa fluid. All the velocities are expressed in units of $\omega_{\rm p}a$ to be consistent with the fits from Ref.~\cite{KalmanPRL2000} and are plotted as functions of the screening parameter $\kappa$. The solid curves correspond to the calculation using Eq.~(\ref{elasticsound_general}) with Eq.~(\ref{longitudinal}) for the longitudinal velocity and Eq.~(\ref{transverse}) for the transverse velocity (these are conventionally referred to as variational). The dashed curves correspond to the fits based on QLCA calculation with RDFs $g(r)$ obtained by following the hypernetted chain scheme as described in Ref.~\cite{KalmanPRL2000}. The kinetic contribution is not included in the instantaneous sound velocities. The dash-dotted curve shows the adiabatic sound velocity.}
\label{Fig2}
\end{figure}

To verify the accuracy of this approximation, we plot in Fig.~\ref{Fig2} the elastic longitudinal and transverse sound velocities as calculated from Eqs. (\ref{elasticsound_general}) - (\ref{transverse}). For comparison, we also plot the approximate expressions for instantaneous elastic sound velocities derived in Ref.~\cite{KalmanPRL2000} within the QLCA framework combined with the hypernetted chain scheme to compute $g(r)$. The sound velocities are expressed in units of $\omega_{\rm p}a$ to be consistent with the fits from Ref.~\cite{KalmanPRL2000}. In addition, in Fig.~\ref{Fig2} we plot the adiabatic sound velocity calculated using a simple practical equation of state of the Yukawa fluid, as discussed in more detail below. At the moment, note how close the instantaneous longitudinal and adiabatic sound velocities are to each other in strong coupling.        

\section{Weak coupling limit} 

The weak coupling limit corresponds to a fully disordered system of Yukawa particles. The HS reference system is not required in this case. We can estimate the properties of the Yukawa fluid in this limit using the fully disordered RDF of the form $g(r)=1$. For excess thermodynamic properties, we get 
\begin{equation}
u_{\rm ex}=p_{\rm ex}=\frac{3\Gamma}{2\kappa^2}.   
\end{equation}
Note that the excess energy and pressure are not necessary much smaller than that of the ideal gas contribution (3/2 for energy and 1 for pressure), especially when $\kappa<1$. This difference from conventional ideal gas is related to the very soft and long-range character of interparticle interaction. Note also that the presence of neutralizing background would cancel the excess energy and pressure of a disordered system of charged particles. For instantaneous elastic sound velocities, we get
\begin{equation}\label{WC_sounds}
c_{l}^2 = 3v_{\rm T}^2+ \omega_p^2\lambda^2, \quad c_{t}^2 = v_{\rm T}^2, \quad c_{b}^2 =\frac{5}{3}v_{\rm T}^2+\omega_p^2\lambda^2.   
\end{equation}
Now, the kinetic contributions are included. Remarkably, the potential contribution to the instantaneous shear modulus and the sound velocity is zero, and only the kinetic contribution survives. Instantaneous longitudinal and bulk sound velocities contain both kinetic and potential contributions. The potential contributions are equal, and they resemble the familiar expression for the dust acoustic wave (DAW) sound velocity. Given the importance of the DAW concept in complex (dusty) plasmas, it makes sense to elaborate on this in more detail.   

First, we recall the canonical derivation of the DAW dispersion relation as per Rao, Shukla and Yu~\cite{Rao1990}.
We adopt the simplest electrostatic description of this phenomenon. The electrons and ions are treated as a polarizable neutralizing medium and are described by
\begin{equation}\label{Boltzmann_ei}
-en_{\rm i}\nabla\varphi = T_{\rm i}\nabla n_{\rm i}, \quad en_{\rm e}\nabla\varphi = T_{\rm e}\nabla n_{\rm e}, 
\end{equation}
where $n_{i,e}$ and $T_{i,e}$ are electron and ion densities and temperatures, while $\varphi$ is the electric potential. This corresponds to the equilibrium Boltzmann response of ion and electron densities to the electric potential of the wave. The charged dust particles are treated as an inertial component and are described by the continuity and momentum equations 
\begin{equation}\label{continuity}
\frac{\partial n}{\partial t}+\nabla(n{\bf v})=0,
\end{equation}
\begin{equation}\label{momentum}
\frac{\partial {\bf v}}{\partial t}+({\bf v}\cdot \nabla){\bf v}=-\frac{Q\nabla\varphi}{m}-\frac{\nabla P}{mn},
\end{equation}
where ${\bf v}$ is the velocity of the particles. In the limit of long-wavelength perturbations (acoustic regime) the system is quasi-neutral and the Poisson equation reduces to
\begin{equation}\label{quasineutrality}
e n_{\rm i} -e n_{\rm e} +Qn = 0.
\end{equation}

The standard linearization procedure is then applied to the set of equations (\ref{Boltzmann_ei})-(\ref{quasineutrality}). We assume $n_{\rm e}=n_{\rm e 0}+n_{\rm e 1}$, $n_{\rm i}=n_{\rm i 0}+n_{\rm i 1}$, $n = n_0+n_1$, $\varphi=\varphi_{1}$, ${\bf v}={\bf v}_1$, and $P=P_0+P_1$, where the quantities with the subscript ``1'' correspond to small perturbations compared to the equilibrium values denoted by the subscript ``0''.  All perturbations are proportional to $\propto\exp(i{\bf kr}-i\omega t)$, where  $\omega$ is the wave frequency, and ${\bf k}$ is the wave vector. For a quasi-neutral disordered system, the pressure contains only the kinetic contribution $P=nT$. Since the sound wave corresponds to an adiabatic perturbation~\cite{LL_Fluids}, the small change in pressure $P_1$ is related to the small change in density $n_1$ by  $P_1=(\partial P/\partial n)_{s}n_1$, where $s$ is the system entropy, which is constant for an adiabatic process. Using the thermodynamic relation $P\propto n^{\gamma}$ with $\gamma=c_{\rm p}/c_{\rm v}$ being the ratio of specific heats (known as the adiabatic index), we get $P_1=\gamma Pn_1/n$. This results in the dispersion relation of the acoustic type with the dust acoustic (DA) velocity
\begin{equation}\label{cDA}
c_{\rm DA}^2\equiv\frac{\omega^2}{k^2}=\gamma v_{\rm T}^2+\omega_{\rm p}^2\lambda^2.
\end{equation}
Comparing Eqs.~(\ref{cDA}) and (\ref{WC_sounds}) we immediately see that for $\gamma = 3$ the DA velocity would coincide with the instantaneous longitudinal elastic sound speed, while for $\gamma = 5/3$ it would coincide with the instantaneous bulk elastic sound speed. Unfortunately, a simple electrostatic description cannot specify the exact value of $\gamma$.  

Let us therefore complement this consideration by evaluating the adiabatic sound velocity of a fully disordered one component Yukawa fluid.
The conventional thermodynamic definition reads~\cite{LL_Fluids}
\begin{equation}\label{definition}
c_{\rm s}^2 = {\left(\frac{\partial P}{\partial \rho_m}\right)_S}=\frac{\gamma}{m}\left(\frac{\partial P}{\partial n}\right)_T,
\end{equation}
where $\rho_m=mn$ is the mass density. Expressing the pressure as a superposition of the ideal and excess contributions, this can be further transformed into~\cite{KhrapakPRE2025}
\begin{equation}\label{definition1}
c_{\rm s}^2 = \gamma v_{\rm T}^2\left(1+p_{\rm ex}+n\frac{\partial p_{\rm ex}}{\partial n}\right)= \gamma\mu v_{\rm T}^2,
\end{equation}
where $\mu$ is the isothermal compressibility modulus. As we already know $p_{\rm ex}=3\Gamma/2\kappa^2$ in the fully disordered limit. This implies that the isothermal compressibility modulus
becomes (see, e.g., Ref.~\cite{KhrapakPRE03_2015})
\begin{equation}
 \mu=1+p_{\rm ex}+\frac{\Gamma}{3}\frac{\partial p_{\rm ex}}{\partial \Gamma}- \frac{\kappa}{3}\frac{\partial p_{\rm ex}}{\partial \kappa} = 1+\frac{3\Gamma}{\kappa^2}.
 \end{equation}
The specific heat at constant volume is calculated from $c_{\rm v}= 3/2+ u_{\rm ex}-\Gamma(\partial u_{\rm ex}/\partial \Gamma)=3/2$. The adiabatic index is evaluated from~\cite{KhrapakPRE03_2015}
\begin{equation}
 \gamma=1+ \frac{[1+p_{\rm ex}-\Gamma(\partial p_{\rm ex}/\partial \Gamma)]^2}{c_{\rm v}\mu}=1+\frac{2}{3\left(1+\frac{3\Gamma}{\kappa^2}\right)}.  
\end{equation}
We obtain for the adiabatic sound velocity
\begin{equation}\label{sound_disordered}
c_{\rm s}^2 = v_{\rm T}^2\left(\frac53+\frac{3\Gamma}{\kappa^2}\right)=\frac53  v_{\rm T}^2+\omega_{\rm p}^2\lambda^2.
\end{equation}
This result suggests that $\gamma=5/3$ is an appropriate adiabatic index for the DA velocity in the weakly coupled limit, similar to conventional dilute neutral gases.   

In the field of dusty plasma, the kinetic term in Eq.~(\ref{cDA}) is often neglected (for instance, in the original derivation Rao, Shukla and Yu assumed that the dust particles are ``cold``~\cite{Rao1990}). The dust-acoustic velocity becomes
\begin{equation}\label{DA_standard}
c_{\rm DA} = \omega_{\rm p}\lambda\simeq \sqrt{Z^2\frac{T_i}{m}\frac{n}{n_i}},
\end{equation}
where $Z=|Q|/e$ is the particle charge number and $e$ is the elementary charge~\cite{BarkanPoP1995}. It is assumed here that screening is mostly associated with the ion component and therefore $\lambda$ can be approximated by the ion Debye radius $\lambda\simeq\lambda_{{\rm D}i}=\sqrt{T_i/4\pi e^2 n_i}$. The dust-acoustic velocity given by Eq.~(\ref{DA_standard}) is frequently assumed to be a universal property of a dusty plasma. Then it can be routinely used to estimate the particle charge number experimentally by measuring the phase velocity of the low-frequency dust wave and simultaneously measuring or making reasonable assumptions about other plasma parameters: $T_i$, $n$, and $n_i$~\cite{ThompsonPoP1997,KhrapakPoP01_2003,AnnibaldiNJP2007,SchwabeEPL2011}.

However, it is important to stress that the dust acoustic velocity not universal. First, it emerges as the result of Yukawa interaction between the charged particles in a plasma environment. Should the interaction potential deviate from the Yukawa form (e.g. due to plasma absorption or emission by the particles, ionization in the surrounding plasma, ion wake-mediated interactions, etc.), this would affect the dispersion relation and, consequently, the magnitude of the effective sound velocity~\cite{DelzannoPRL2005,KhrapakPoP02_2017,SchwabeNJP2020}. In addition, the dispersion relation of low-frequency dust density waves is often affected by ion drifts and collisional effects (mainly dust-neutral and ion-neutral collisions), and this can also affect the sound velocity~\cite{dAngeloPSS1996,RosenbergJVST1996,IvlevPoP1999,FortovPoP2000,RatynskaiaPRL2004,PielPRE2008,YaroshenkoPoP2019,KhrapakPPCF2020,BajajPRE2022}. Even in an idealized situation leading to Eq.~(\ref{sound_disordered}), neglecting the kinetic contribution requires $\Gamma\gg 5\kappa^2/9$. This inequality can be inconsistent with the assumption of a fully disordered state of the Yukawa fluid used in the derivation, particularly for finite $\kappa$ values.  In general, strong coupling and particle-particle correlations are known to reduce the sound velocity compared to the $c_{\rm DA}$ scale, and these effects become more and more pronounced as $\kappa$ increases. This has been clearly shown in Refs.~\cite{KhrapakPRE03_2015,KhrapakPPCF2015} and will also become evident in Sec.~\ref{Transition}.

\section{Transitional regime}\label{Transition}

The adiabatic sound velocity characterizes longitudinal sound propagation in the limit of low frequency and long wavelength. At high frequencies, sound propagation is characterized by the instantaneous longitudinal modulus within the concept of viscoelasticity~\cite{AliottaPRE2011,KawPoP2001}. For a fully disordered state, the adiabatic sound velocity coincides with the instantaneous elastic bulk sound velocity. In the strong coupling limit, the kinetic contributions are negligible and the adiabatic sound velocity is close to the instantaneous elastic longitudinal and bulk sound velocities~\cite{KhrapakPoP02_2016_1,KhrapakPoF2023}. This is related to the fact that for a soft and long-ranged Yukawa potential with not too large $\kappa$, the strong inequality $c_l\gg c_t$ holds~\cite{KhrapakPoP10_2019}. Combined with the relation $c_l^2=c_b^2+(4/3)c_t^2$, this explains the observation. The natural next question is how the adiabatic and instantaneous elastic sound velocities are related in the transitional regime.   

In fact, there is no full consensus on what controls the sound velocity in the transitional regime. Based on the results obtained in the weak coupling limit, it seems natural to assume $c_{\rm s}\simeq c_{\rm b}$. However, other assumptions have also been made in the literature. For example, in Ref.~\cite{KhrapakPRE09_2020} based on a thorough comparison between different theoretical approaches and MD simulations it was concluded that, in the weak and moderate coupling regimes, the best description is provided by the sum of the excess bulk modulus and the Bohm-Gross kinetic term. This intuitive phenomenological approach was then supported by the results of the static local field correction formalism~\cite{ToliasPoP2021}. However, the main attention was focused on the dispersion relations of the longitudinal collective mode at finite wave vectors, not particularly on the long-wavelength limit and the sound velocity. 

\begin{figure*}
\includegraphics[width=17cm]{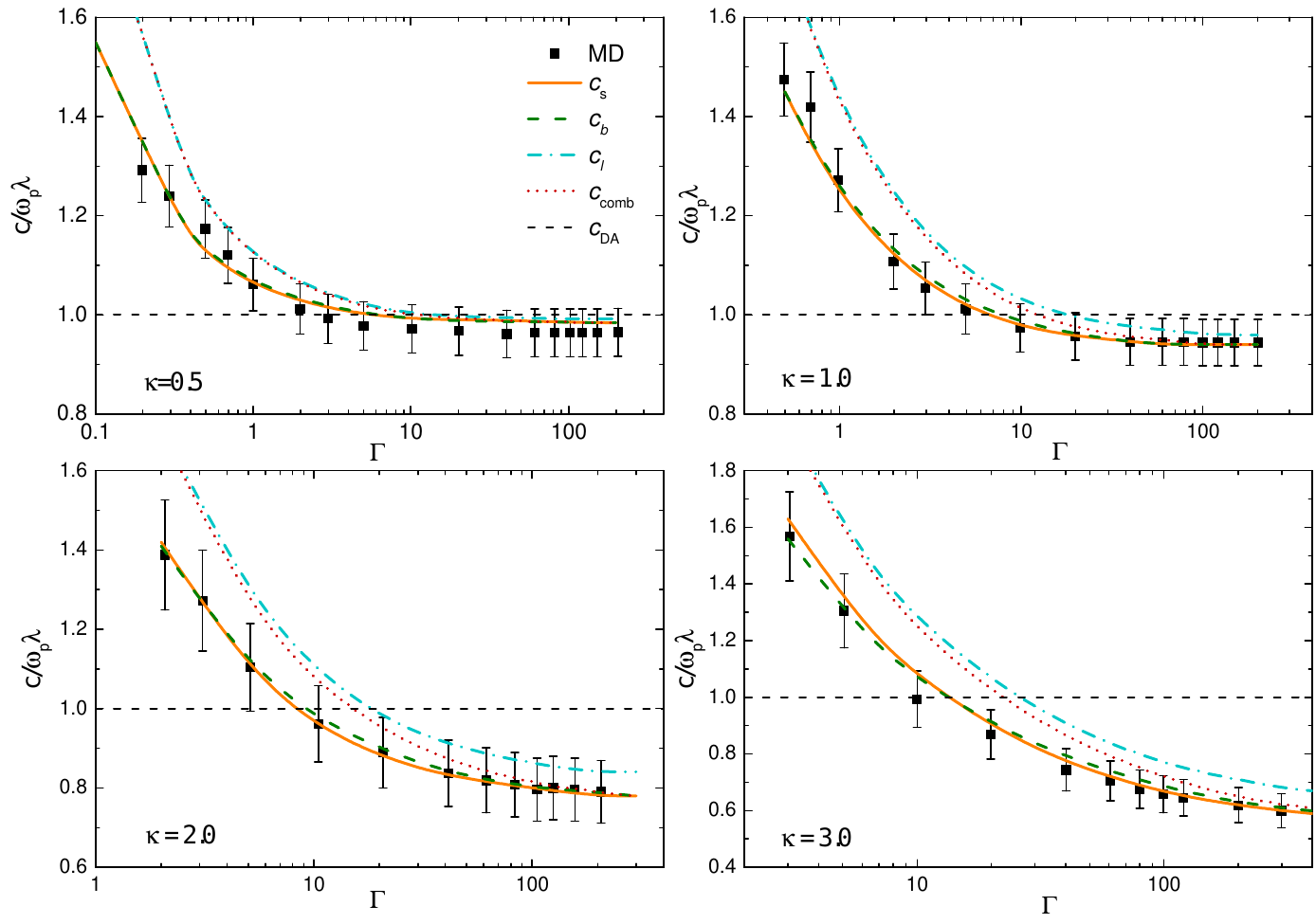}
\caption{(Color online) The reduced sound velocities of the Yukawa fluid across coupling regimes for four values of the screening parameter, $\kappa = 0.5$, 1.0, 2.0, and 3.0 (indicated in the panels). All the velocities are expressed in units of $\omega_{\rm p}\lambda$ to be consistent with the MD results from Ref.~\cite{SilvestriPRE2019} and are shown as functions of the coupling parameter $\Gamma$. The black squares with 5\% error bars at $\kappa = 0.5$ and 1.0 and 10\% error bars at $\kappa = 2.0$ and 3.0 correspond to MD simulations performed in Ref.~\cite{SilvestriPRE2019}. The solid curves correspond to the adiabatic sound velocity $c_{\rm s}$ evaluated using the practical formulas for the excess energy and pressure of Yukawa fluids~\cite{KhrapakPRE02_2015,KhrapakPRE03_2015}. The dashed and dash-dotted curves are the sound velocities corresponding to the instantaneous bulk ($c_{\rm b}$) and longitudinal ($c_{\rm l}$) elastic moduli with kinetic contributions retained. The short-dashed curve corresponds to an empirical expression ($c_{\rm comb}$). The horizontal dashed line at $c/\omega_{\rm p}\lambda=1$ in each panel corresponds to the conventional dust-acoustic velocity $c_{\rm DA}$.}
\label{Fig3}
\end{figure*}

Recently, the evolution of the sound velocity in the one-component Yukawa fluid from the weak to strong-coupling regimes was studied by analyzing the dynamic structure factor $S({\bf k},\omega)$ obtained from MD computer simulation in the low-frequency domain~\cite{SilvestriPRE2019}.
A large body of high-quality data have been obtained in a rather wide range of $\Gamma$ and $\kappa$ values. This data set provides an excellent opportunity to assess which theoretical models offer the most cogent physical description and the best agreement with MD data across coupling regimes. 

Several theoretical models have already been tested~\cite{SilvestriPRE2019}. The results of these tests are shown in Figs.~7 and 8 in Ref. ~\cite{SilvestriPRE2019}. A short summary is that the adiabatic sound velocity calculated using the practical equation of state proposed by Khrapak and Thomas~\cite{KhrapakPRE02_2015,KhrapakPRE03_2015} is in very good agreement with the MD results across the coupling and screening regimes. The random phase approximation and the collisional thermodynamic approaches with the adiabatic factor $\gamma = 5/3$ perform relatively well at weak coupling and are particularly appropriate at weak screening. The QLCA approach demonstrates high accuracy in the strong coupling regime, as expected. The sum-rule-based approach (Feynman ansatz) somewhat underestimates the MD results. We do not repeat this comparison here, but we concentrate on instantaneous elastic sound velocities that can be calculated using the variational approach instead. 

Figure~\ref{Fig3} shows the comparison. The solid squares with error bars are the MD results from Ref.~\cite{SilvestriPRE2019}. The solid curves show the calculation based on the thermodynamic definition of Eq.~(\ref{definition}) using practical EoS with RT scaling~\cite{KhrapakPRE02_2015,KhrapakPRE03_2015}. Although the variational approach provides direct access to the excess pressure and energy, calculation of the sound velocity involves derivatives of these quantities with respect to $\Gamma$ and $\kappa$~\cite{KhrapakPRE03_2015}. Parametric form makes these derivatives non-trivial and we prefer practical expressions proposed in Ref.~\cite{KhrapakPRE03_2015}.
The formulas contain two parameters $\delta$ and $\epsilon$ which enter the RT expression for the thermal component of excess energy $u_{\rm th}=\delta(\Gamma/\Gamma_{\rm m})^{2/5}+\epsilon$; here we use  $\delta=3.0$ and $\epsilon=0$ as suggested by Rosenfeld~\cite{RosenfeldPRE2000}. As expected and demonstrated previously in Ref.~\cite{SilvestriPRE2019}, the adiabatic sound velocity is in excellent agreement with the MD data across coupling regimes. The dashed curves show the instantaneous bulk elastic sound velocity $c_{\rm b}$ with the ideal gas contribution retained. They are also in excellent agreement with the MD data and almost ideally coincide with $c_{\rm s}$. The dashed-dotted curves represent the instantaneous longitudinal elastic sound speed, $c_l$. These curves overestimate the MD data in the weak coupling regime because the ideal gas contribution is overestimated. At strong coupling, the longitudinal sound speed approaches the MD results, but still slightly overestimates them. The short-dashed curves correspond to the phenomenological expression $c_{\rm comb}^2=\tfrac43v_{\rm T}^2+c_b^2$, which combines the configurational contribution to the instantaneous bulk modulus with the Bohm-Gross kinetic term. This model also overestimates the MD data in weak coupling and agrees reasonably well with the MD data in strong coupling (better than the longitudinal sound speed $c_l$). The horizontal dashed line corresponds to the DA velocity scale (\ref{DA_standard}). We see that this model only provides a reasonable description of the actual sound speed in strong coupling and weak screening ($\kappa=0.5$). At weak coupling, it is inaccurate because the kinetic contribution is not included. At strong coupling, particle-particle correlations play an increasingly important role as $\kappa$ increases. This is the reason why the actual sound velocity drops below the DA scale.    

The demonstrated agreement between $c_{\rm s}$ and  $c_{b}$ is striking. It is important to note that this is not a general relationship. In Ref.~\cite{KhrapakPoF2023} we have demonstrated that although it holds for repulsive potentials (inverse power law and Yukawa) in the soft interaction limit, it may be violated in other regimes. For example, in the Lennard-Jones fluid $c_{\rm s}$ and $c_{b}$ can differ considerably~\cite{KhrapakPoF2023}.

\section{Conclusion}
 
We have demonstrated that a variational method based on the Bogoliubov inequality gives access not only to thermodynamic properties of multi-particle systems, such as internal and free energies, but also to other properties expressed with the help of integrals involving the RDF. Here we implemented the variational method to calculate the instantaneous elastic moduli of the Yukawa fluid, using the hard-sphere fluid as a reference system. Our results are in excellent agreement with those obtained using more sophisticated approaches. We have discussed in detail the effects that control the sound velocity of the Yukawa fluid in the weak- and strong-coupling regimes. The relations between the adiabatic sound speed and various instantaneous elastic sound speeds across coupling regimes have been clarified. From practical perspective, these results can be helpful when dealing with elastic properties of strongly coupled Yukawa fluids. The topics can range from dispersion relations and instability conditions of low-frequency particle modes, estimation of particle charge from experimentally measured sound velocities, to identification of appropriate relaxation times.



\bibliography{SE_Ref}

\providecommand{\noopsort}[1]{}\providecommand{\singleletter}[1]{#1}%
\begin{thebibliography}{90}%
\makeatletter
\providecommand \@ifxundefined [1]{%
 \@ifx{#1\undefined}
}%
\providecommand \@ifnum [1]{%
 \ifnum #1\expandafter \@firstoftwo
 \else \expandafter \@secondoftwo
 \fi
}%
\providecommand \@ifx [1]{%
 \ifx #1\expandafter \@firstoftwo
 \else \expandafter \@secondoftwo
 \fi
}%
\providecommand \natexlab [1]{#1}%
\providecommand \enquote  [1]{``#1''}%
\providecommand \bibnamefont  [1]{#1}%
\providecommand \bibfnamefont [1]{#1}%
\providecommand \citenamefont [1]{#1}%
\providecommand \href@noop [0]{\@secondoftwo}%
\providecommand \href [0]{\begingroup \@sanitize@url \@href}%
\providecommand \@href[1]{\@@startlink{#1}\@@href}%
\providecommand \@@href[1]{\endgroup#1\@@endlink}%
\providecommand \@sanitize@url [0]{\catcode `\\12\catcode `\$12\catcode
  `\&12\catcode `\#12\catcode `\^12\catcode `\_12\catcode `\%12\relax}%
\providecommand \@@startlink[1]{}%
\providecommand \@@endlink[0]{}%
\providecommand \url  [0]{\begingroup\@sanitize@url \@url }%
\providecommand \@url [1]{\endgroup\@href {#1}{\urlprefix }}%
\providecommand \urlprefix  [0]{URL }%
\providecommand \Eprint [0]{\href }%
\providecommand \doibase [0]{https://doi.org/}%
\providecommand \selectlanguage [0]{\@gobble}%
\providecommand \bibinfo  [0]{\@secondoftwo}%
\providecommand \bibfield  [0]{\@secondoftwo}%
\providecommand \translation [1]{[#1]}%
\providecommand \BibitemOpen [0]{}%
\providecommand \bibitemStop [0]{}%
\providecommand \bibitemNoStop [0]{.\EOS\space}%
\providecommand \EOS [0]{\spacefactor3000\relax}%
\providecommand \BibitemShut  [1]{\csname bibitem#1\endcsname}%
\let\auto@bib@innerbib\@empty
\bibitem [{\citenamefont {Zwanzig}\ and\ \citenamefont
  {Mountain}(1965)}]{ZwanzigJCP1965}%
  \BibitemOpen
  \bibfield  {author} {\bibinfo {author} {\bibfnamefont {R.}~\bibnamefont
  {Zwanzig}}\ and\ \bibinfo {author} {\bibfnamefont {R.~D.}\ \bibnamefont
  {Mountain}},\ }\bibfield  {title} {\bibinfo {title} {High-frequency elastic
  moduli of simple fluids},\ }\href {https://doi.org/10.1063/1.1696718}
  {\bibfield  {journal} {\bibinfo  {journal} {J. Chem. Phys.}\ }\textbf
  {\bibinfo {volume} {43}},\ \bibinfo {pages} {4464} (\bibinfo {year}
  {1965})}\BibitemShut {NoStop}%
\bibitem [{\citenamefont {Hubbard}\ and\ \citenamefont
  {Beeby}(1969)}]{Hubbard1969}%
  \BibitemOpen
  \bibfield  {author} {\bibinfo {author} {\bibfnamefont {J.}~\bibnamefont
  {Hubbard}}\ and\ \bibinfo {author} {\bibfnamefont {J.~L.}\ \bibnamefont
  {Beeby}},\ }\bibfield  {title} {\bibinfo {title} {Collective motion in
  liquids},\ }\href {https://doi.org/10.1088/0022-3719/2/3/318} {\bibfield
  {journal} {\bibinfo  {journal} {J. Phys. C}\ }\textbf {\bibinfo {volume}
  {2}},\ \bibinfo {pages} {556} (\bibinfo {year} {1969})}\BibitemShut {NoStop}%
\bibitem [{\citenamefont {Balucani}\ and\ \citenamefont
  {Zoppi}(1994)}]{BalucaniBook}%
  \BibitemOpen
  \bibfield  {author} {\bibinfo {author} {\bibfnamefont {U.}~\bibnamefont
  {Balucani}}\ and\ \bibinfo {author} {\bibfnamefont {M.}~\bibnamefont
  {Zoppi}},\ }\href@noop {} {\emph {\bibinfo {title} {Dynamics of the Liquid
  State}}}\ (\bibinfo  {publisher} {Clarendon Press},\ \bibinfo {address}
  {Oxford},\ \bibinfo {year} {1994})\BibitemShut {NoStop}%
\bibitem [{\citenamefont {Morkel}\ \emph {et~al.}(1993)\citenamefont {Morkel},
  \citenamefont {Bodensteiner},\ and\ \citenamefont
  {Gemperlein}}]{MorkelPRE1993}%
  \BibitemOpen
  \bibfield  {author} {\bibinfo {author} {\bibfnamefont {C.}~\bibnamefont
  {Morkel}}, \bibinfo {author} {\bibfnamefont {T.}~\bibnamefont
  {Bodensteiner}},\ and\ \bibinfo {author} {\bibfnamefont {H.}~\bibnamefont
  {Gemperlein}},\ }\bibfield  {title} {\bibinfo {title} {Zero-sound-like modes
  in simple liquid metals},\ }\href {https://doi.org/10.1103/physreve.47.2575}
  {\bibfield  {journal} {\bibinfo  {journal} {Phys. Rev. E}\ }\textbf {\bibinfo
  {volume} {47}},\ \bibinfo {pages} {2575} (\bibinfo {year}
  {1993})}\BibitemShut {NoStop}%
\bibitem [{\citenamefont {Golden}\ and\ \citenamefont
  {Kalman}(2000)}]{GoldenPoP2000}%
  \BibitemOpen
  \bibfield  {author} {\bibinfo {author} {\bibfnamefont {K.~I.}\ \bibnamefont
  {Golden}}\ and\ \bibinfo {author} {\bibfnamefont {G.~J.}\ \bibnamefont
  {Kalman}},\ }\bibfield  {title} {\bibinfo {title} {Quasilocalized charge
  approximation in strongly coupled plasma physics},\ }\href
  {https://doi.org/10.1063/1.873814} {\bibfield  {journal} {\bibinfo  {journal}
  {Phys. Plasmas}\ }\textbf {\bibinfo {volume} {7}},\ \bibinfo {pages} {14}
  (\bibinfo {year} {2000})}\BibitemShut {NoStop}%
\bibitem [{\citenamefont {Kaw}(2001)}]{KawPoP2001}%
  \BibitemOpen
  \bibfield  {author} {\bibinfo {author} {\bibfnamefont {P.~K.}\ \bibnamefont
  {Kaw}},\ }\bibfield  {title} {\bibinfo {title} {Collective modes in a
  strongly coupled dusty plasma},\ }\href {https://doi.org/10.1063/1.1348335}
  {\bibfield  {journal} {\bibinfo  {journal} {Phys. Plasmas}\ }\textbf
  {\bibinfo {volume} {8}},\ \bibinfo {pages} {1870} (\bibinfo {year}
  {2001})}\BibitemShut {NoStop}%
\bibitem [{\citenamefont {Aliotta}\ \emph {et~al.}(2011)\citenamefont
  {Aliotta}, \citenamefont {Gapiński}, \citenamefont {Pochylski},
  \citenamefont {Ponterio}, \citenamefont {Saija},\ and\ \citenamefont
  {Vasi}}]{AliottaPRE2011}%
  \BibitemOpen
  \bibfield  {author} {\bibinfo {author} {\bibfnamefont {F.}~\bibnamefont
  {Aliotta}}, \bibinfo {author} {\bibfnamefont {J.}~\bibnamefont {Gapiński}},
  \bibinfo {author} {\bibfnamefont {M.}~\bibnamefont {Pochylski}}, \bibinfo
  {author} {\bibfnamefont {R.~C.}\ \bibnamefont {Ponterio}}, \bibinfo {author}
  {\bibfnamefont {F.}~\bibnamefont {Saija}},\ and\ \bibinfo {author}
  {\bibfnamefont {C.}~\bibnamefont {Vasi}},\ }\bibfield  {title} {\bibinfo
  {title} {Collective acoustic modes in liquids: {A} comparison between the
  generalized-hydrodynamics and memory-function approaches},\ }\href
  {https://doi.org/10.1103/physreve.84.051202} {\bibfield  {journal} {\bibinfo
  {journal} {Phys. Rev. E}\ }\textbf {\bibinfo {volume} {84}},\ \bibinfo
  {pages} {051202} (\bibinfo {year} {2011})}\BibitemShut {NoStop}%
\bibitem [{\citenamefont {Bryk}\ \emph {et~al.}(2014)\citenamefont {Bryk},
  \citenamefont {Gorelli}, \citenamefont {Ruocco}, \citenamefont {Santoro},\
  and\ \citenamefont {Scopigno}}]{BrykPRE2014}%
  \BibitemOpen
  \bibfield  {author} {\bibinfo {author} {\bibfnamefont {T.}~\bibnamefont
  {Bryk}}, \bibinfo {author} {\bibfnamefont {F.}~\bibnamefont {Gorelli}},
  \bibinfo {author} {\bibfnamefont {G.}~\bibnamefont {Ruocco}}, \bibinfo
  {author} {\bibfnamefont {M.}~\bibnamefont {Santoro}},\ and\ \bibinfo {author}
  {\bibfnamefont {T.}~\bibnamefont {Scopigno}},\ }\bibfield  {title} {\bibinfo
  {title} {Collective excitations in soft-sphere fluids},\ }\href
  {https://doi.org/10.1103/physreve.90.042301} {\bibfield  {journal} {\bibinfo
  {journal} {Phys. Rev. E}\ }\textbf {\bibinfo {volume} {90}},\ \bibinfo
  {pages} {042301} (\bibinfo {year} {2014})}\BibitemShut {NoStop}%
\bibitem [{\citenamefont {Hosokawa}\ \emph {et~al.}(2015)\citenamefont
  {Hosokawa}, \citenamefont {Inui}, \citenamefont {Kajihara}, \citenamefont
  {Tsutsui},\ and\ \citenamefont {Baron}}]{HosokawaJPCM2015}%
  \BibitemOpen
  \bibfield  {author} {\bibinfo {author} {\bibfnamefont {S.}~\bibnamefont
  {Hosokawa}}, \bibinfo {author} {\bibfnamefont {M.}~\bibnamefont {Inui}},
  \bibinfo {author} {\bibfnamefont {Y.}~\bibnamefont {Kajihara}}, \bibinfo
  {author} {\bibfnamefont {S.}~\bibnamefont {Tsutsui}},\ and\ \bibinfo {author}
  {\bibfnamefont {A.~Q.~R.}\ \bibnamefont {Baron}},\ }\bibfield  {title}
  {\bibinfo {title} {Transverse excitations in liquid {Fe}, {Cu} and {Zn}},\
  }\href {https://doi.org/10.1088/0953-8984/27/19/194104} {\bibfield  {journal}
  {\bibinfo  {journal} {J. Phys.: Condens. Matter}\ }\textbf {\bibinfo {volume}
  {27}},\ \bibinfo {pages} {194104} (\bibinfo {year} {2015})}\BibitemShut
  {NoStop}%
\bibitem [{\citenamefont {Khrapak}\ \emph
  {et~al.}(2017{\natexlab{a}})\citenamefont {Khrapak}, \citenamefont {Klumov},\
  and\ \citenamefont {Couedel}}]{KhrapakSciRep2017}%
  \BibitemOpen
  \bibfield  {author} {\bibinfo {author} {\bibfnamefont {S.}~\bibnamefont
  {Khrapak}}, \bibinfo {author} {\bibfnamefont {B.}~\bibnamefont {Klumov}},\
  and\ \bibinfo {author} {\bibfnamefont {L.}~\bibnamefont {Couedel}},\
  }\bibfield  {title} {\bibinfo {title} {Collective modes in simple melts:
  Transition from soft spheres to the hard sphere limit},\ }\href
  {https://doi.org/10.1038/s41598-017-08429-5} {\bibfield  {journal} {\bibinfo
  {journal} {Sci. Rep.}\ }\textbf {\bibinfo {volume} {7}},\ \bibinfo {pages}
  {7985} (\bibinfo {year} {2017}{\natexlab{a}})}\BibitemShut {NoStop}%
\bibitem [{\citenamefont {Khrapak}\ \emph {et~al.}(2021)\citenamefont
  {Khrapak}, \citenamefont {Kryuchkov}, \citenamefont {Mistryukova},\ and\
  \citenamefont {Yurchenko}}]{KhrapakPRE05_2021}%
  \BibitemOpen
  \bibfield  {author} {\bibinfo {author} {\bibfnamefont {S.}~\bibnamefont
  {Khrapak}}, \bibinfo {author} {\bibfnamefont {N.~P.}\ \bibnamefont
  {Kryuchkov}}, \bibinfo {author} {\bibfnamefont {L.~A.}\ \bibnamefont
  {Mistryukova}},\ and\ \bibinfo {author} {\bibfnamefont {S.~O.}\ \bibnamefont
  {Yurchenko}},\ }\bibfield  {title} {\bibinfo {title} {From soft- to
  hard-sphere fluids: Crossover evidenced by high-frequency elastic moduli},\
  }\href {https://doi.org/10.1103/physreve.103.052117} {\bibfield  {journal}
  {\bibinfo  {journal} {Phys. Rev. E}\ }\textbf {\bibinfo {volume} {103}},\
  \bibinfo {pages} {052117} (\bibinfo {year} {2021})}\BibitemShut {NoStop}%
\bibitem [{\citenamefont {Mountain}\ and\ \citenamefont
  {Zwanzig}(1966)}]{MountainJCP1966}%
  \BibitemOpen
  \bibfield  {author} {\bibinfo {author} {\bibfnamefont {R.~D.}\ \bibnamefont
  {Mountain}}\ and\ \bibinfo {author} {\bibfnamefont {R.}~\bibnamefont
  {Zwanzig}},\ }\bibfield  {title} {\bibinfo {title} {Shear relaxation times of
  simple fluids},\ }\href {https://doi.org/10.1063/1.1727124} {\bibfield
  {journal} {\bibinfo  {journal} {J. Chem. Phys.}\ }\textbf {\bibinfo {volume}
  {44}},\ \bibinfo {pages} {2777–2779} (\bibinfo {year} {1966})}\BibitemShut
  {NoStop}%
\bibitem [{\citenamefont {Trachenko}(2023)}]{TrachenkoBook}%
  \BibitemOpen
  \bibfield  {author} {\bibinfo {author} {\bibfnamefont {K.}~\bibnamefont
  {Trachenko}},\ }\href@noop {} {\emph {\bibinfo {title} {Theory of liquids:
  {F}rom excitations to Thermodynamics}}}\ (\bibinfo  {publisher} {Cambridge
  University Press},\ \bibinfo {address} {Cambridge, England},\ \bibinfo {year}
  {2023})\BibitemShut {NoStop}%
\bibitem [{\citenamefont {Hartkamp}\ \emph {et~al.}(2013)\citenamefont
  {Hartkamp}, \citenamefont {Daivis},\ and\ \citenamefont
  {Todd}}]{HartkampPRE2013}%
  \BibitemOpen
  \bibfield  {author} {\bibinfo {author} {\bibfnamefont {R.}~\bibnamefont
  {Hartkamp}}, \bibinfo {author} {\bibfnamefont {P.~J.}\ \bibnamefont
  {Daivis}},\ and\ \bibinfo {author} {\bibfnamefont {B.~D.}\ \bibnamefont
  {Todd}},\ }\bibfield  {title} {\bibinfo {title} {Density dependence of the
  stress relaxation function of a simple fluid},\ }\href
  {https://doi.org/10.1103/physreve.87.032155} {\bibfield  {journal} {\bibinfo
  {journal} {Phys. Rev. E}\ }\textbf {\bibinfo {volume} {87}},\ \bibinfo
  {pages} {032155} (\bibinfo {year} {2013})}\BibitemShut {NoStop}%
\bibitem [{\citenamefont {Khrapak}\ and\ \citenamefont
  {Khrapak}(2024)}]{KhrapakPRE11_2024}%
  \BibitemOpen
  \bibfield  {author} {\bibinfo {author} {\bibfnamefont {S.~A.}\ \bibnamefont
  {Khrapak}}\ and\ \bibinfo {author} {\bibfnamefont {A.~G.}\ \bibnamefont
  {Khrapak}},\ }\bibfield  {title} {\bibinfo {title} {Quasiuniversal behavior
  of shear relaxation times in simple fluids},\ }\href
  {https://doi.org/10.1103/physreve.110.054101} {\bibfield  {journal} {\bibinfo
   {journal} {Phys. Rev. E}\ }\textbf {\bibinfo {volume} {110}},\ \bibinfo
  {pages} {054101} (\bibinfo {year} {2024})}\BibitemShut {NoStop}%
\bibitem [{\citenamefont {Khrapak}(2021{\natexlab{a}})}]{KhrapakPRE01_2021}%
  \BibitemOpen
  \bibfield  {author} {\bibinfo {author} {\bibfnamefont {S.~A.}\ \bibnamefont
  {Khrapak}},\ }\bibfield  {title} {\bibinfo {title} {Vibrational model of
  thermal conduction for fluids with soft interactions},\ }\href
  {https://doi.org/10.1103/physreve.103.013207} {\bibfield  {journal} {\bibinfo
   {journal} {Phys. Rev. E}\ }\textbf {\bibinfo {volume} {103}},\ \bibinfo
  {pages} {013207} (\bibinfo {year} {2021}{\natexlab{a}})}\BibitemShut
  {NoStop}%
\bibitem [{\citenamefont {Khrapak}(2021{\natexlab{b}})}]{KhrapakPoP08_2021}%
  \BibitemOpen
  \bibfield  {author} {\bibinfo {author} {\bibfnamefont {S.~A.}\ \bibnamefont
  {Khrapak}},\ }\bibfield  {title} {\bibinfo {title} {Thermal conductivity of
  strongly coupled {Y}ukawa fluids},\ }\href
  {https://doi.org/10.1063/5.0056763} {\bibfield  {journal} {\bibinfo
  {journal} {Phys. Plasmas}\ }\textbf {\bibinfo {volume} {28}},\ \bibinfo
  {pages} {084501} (\bibinfo {year} {2021}{\natexlab{b}})}\BibitemShut
  {NoStop}%
\bibitem [{\citenamefont
  {Khrapak}(2021{\natexlab{c}})}]{KhrapakMolecules12_2021}%
  \BibitemOpen
  \bibfield  {author} {\bibinfo {author} {\bibfnamefont {S.~A.}\ \bibnamefont
  {Khrapak}},\ }\bibfield  {title} {\bibinfo {title} {Self-diffusion in simple
  liquids as a random walk process},\ }\href
  {https://doi.org/10.3390/molecules26247499} {\bibfield  {journal} {\bibinfo
  {journal} {Molecules}\ }\textbf {\bibinfo {volume} {26}},\ \bibinfo {pages}
  {7499} (\bibinfo {year} {2021}{\natexlab{c}})}\BibitemShut {NoStop}%
\bibitem [{\citenamefont {Khrapak}\ and\ \citenamefont
  {Khrapak}(2023{\natexlab{a}})}]{KhrapakPRE12_2023}%
  \BibitemOpen
  \bibfield  {author} {\bibinfo {author} {\bibfnamefont {S.~A.}\ \bibnamefont
  {Khrapak}}\ and\ \bibinfo {author} {\bibfnamefont {A.~G.}\ \bibnamefont
  {Khrapak}},\ }\bibfield  {title} {\bibinfo {title} {Vibrational model for
  thermal conductivity of {L}ennard-{J}ones fluids: {A}pplicability domain and
  accuracy level},\ }\href {https://doi.org/10.1103/physreve.108.064129}
  {\bibfield  {journal} {\bibinfo  {journal} {Phy. Rev. E}\ }\textbf {\bibinfo
  {volume} {108}},\ \bibinfo {pages} {064129} (\bibinfo {year}
  {2023}{\natexlab{a}})}\BibitemShut {NoStop}%
\bibitem [{\citenamefont {Khrapak}(2024{\natexlab{a}})}]{KhrapakPhysRep2024}%
  \BibitemOpen
  \bibfield  {author} {\bibinfo {author} {\bibfnamefont {S.}~\bibnamefont
  {Khrapak}},\ }\bibfield  {title} {\bibinfo {title} {Elementary vibrational
  model for transport properties of dense fluids},\ }\href
  {https://doi.org/10.1016/j.physrep.2023.11.004} {\bibfield  {journal}
  {\bibinfo  {journal} {Phys. Rep.}\ }\textbf {\bibinfo {volume} {1050}},\
  \bibinfo {pages} {1} (\bibinfo {year} {2024}{\natexlab{a}})}\BibitemShut
  {NoStop}%
\bibitem [{\citenamefont {Zwanzig}(1983)}]{ZwanzigJCP1983}%
  \BibitemOpen
  \bibfield  {author} {\bibinfo {author} {\bibfnamefont {R.}~\bibnamefont
  {Zwanzig}},\ }\bibfield  {title} {\bibinfo {title} {On the relation between
  self-diffusion and viscosity of liquids},\ }\href
  {https://doi.org/10.1063/1.446338} {\bibfield  {journal} {\bibinfo  {journal}
  {J. Chem. Phys.}\ }\textbf {\bibinfo {volume} {79}},\ \bibinfo {pages} {4507}
  (\bibinfo {year} {1983})}\BibitemShut {NoStop}%
\bibitem [{\citenamefont {Khrapak}(2019{\natexlab{a}})}]{KhrapakMolPhys2019}%
  \BibitemOpen
  \bibfield  {author} {\bibinfo {author} {\bibfnamefont {S.}~\bibnamefont
  {Khrapak}},\ }\bibfield  {title} {\bibinfo {title}
  {Stokes{\textendash}{E}instein relation in simple fluids revisited},\ }\href
  {https://doi.org/10.1080/00268976.2019.1643045} {\bibfield  {journal}
  {\bibinfo  {journal} {Mol. Phys.}\ }\textbf {\bibinfo {volume} {118}},\
  \bibinfo {pages} {e1643045} (\bibinfo {year}
  {2019}{\natexlab{a}})}\BibitemShut {NoStop}%
\bibitem [{\citenamefont {Khrapak}\ and\ \citenamefont
  {Khrapak}(2021)}]{KhrapakPRE10_2021}%
  \BibitemOpen
  \bibfield  {author} {\bibinfo {author} {\bibfnamefont {S.~A.}\ \bibnamefont
  {Khrapak}}\ and\ \bibinfo {author} {\bibfnamefont {A.~G.}\ \bibnamefont
  {Khrapak}},\ }\bibfield  {title} {\bibinfo {title} {Excess entropy and
  {S}tokes-{E}instein relation in simple fluids},\ }\href
  {https://doi.org/10.1103/physreve.104.044110} {\bibfield  {journal} {\bibinfo
   {journal} {Phys. Rev. E}\ }\textbf {\bibinfo {volume} {104}},\ \bibinfo
  {pages} {044110} (\bibinfo {year} {2021})}\BibitemShut {NoStop}%
\bibitem [{\citenamefont {Khrapak}\ and\ \citenamefont
  {Yurchenko}(2021)}]{KhrapakJCP2021}%
  \BibitemOpen
  \bibfield  {author} {\bibinfo {author} {\bibfnamefont {S.~A.}\ \bibnamefont
  {Khrapak}}\ and\ \bibinfo {author} {\bibfnamefont {S.~O.}\ \bibnamefont
  {Yurchenko}},\ }\bibfield  {title} {\bibinfo {title} {Entropy of simple
  fluids with repulsive interactions near freezing},\ }\href
  {https://doi.org/10.1063/5.0063559} {\bibfield  {journal} {\bibinfo
  {journal} {J. Chem. Phys.}\ }\textbf {\bibinfo {volume} {155}},\ \bibinfo
  {pages} {134501} (\bibinfo {year} {2021})}\BibitemShut {NoStop}%
\bibitem [{\citenamefont {Khrapak}(2024{\natexlab{b}})}]{KhrapakPRE09_2024}%
  \BibitemOpen
  \bibfield  {author} {\bibinfo {author} {\bibfnamefont {S.~A.}\ \bibnamefont
  {Khrapak}},\ }\bibfield  {title} {\bibinfo {title} {Entropy of strongly
  coupled {Y}ukawa fluids},\ }\href
  {https://doi.org/10.1103/physreve.110.034602} {\bibfield  {journal} {\bibinfo
   {journal} {Phys. Rev. E}\ }\textbf {\bibinfo {volume} {110}},\ \bibinfo
  {pages} {034602} (\bibinfo {year} {2024}{\natexlab{b}})}\BibitemShut
  {NoStop}%
\bibitem [{\citenamefont {Huang}\ \emph {et~al.}(2023)\citenamefont {Huang},
  \citenamefont {Baggioli}, \citenamefont {Lu}, \citenamefont {Ma},\ and\
  \citenamefont {Feng}}]{HuangPRR2023}%
  \BibitemOpen
  \bibfield  {author} {\bibinfo {author} {\bibfnamefont {D.}~\bibnamefont
  {Huang}}, \bibinfo {author} {\bibfnamefont {M.}~\bibnamefont {Baggioli}},
  \bibinfo {author} {\bibfnamefont {S.}~\bibnamefont {Lu}}, \bibinfo {author}
  {\bibfnamefont {Z.}~\bibnamefont {Ma}},\ and\ \bibinfo {author}
  {\bibfnamefont {Y.}~\bibnamefont {Feng}},\ }\bibfield  {title} {\bibinfo
  {title} {Revealing the supercritical dynamics of dusty plasmas and their
  liquidlike to gaslike dynamical crossover},\ }\href
  {https://doi.org/10.1103/physrevresearch.5.013149} {\bibfield  {journal}
  {\bibinfo  {journal} {Phys. Rev. Research}\ }\textbf {\bibinfo {volume}
  {5}},\ \bibinfo {pages} {013149} (\bibinfo {year} {2023})}\BibitemShut
  {NoStop}%
\bibitem [{\citenamefont {Yu}\ \emph {et~al.}(2024)\citenamefont {Yu},
  \citenamefont {Huang}, \citenamefont {Lu}, \citenamefont {Khrapak},\ and\
  \citenamefont {Feng}}]{YuPRE2024}%
  \BibitemOpen
  \bibfield  {author} {\bibinfo {author} {\bibfnamefont {N.}~\bibnamefont
  {Yu}}, \bibinfo {author} {\bibfnamefont {D.}~\bibnamefont {Huang}}, \bibinfo
  {author} {\bibfnamefont {S.}~\bibnamefont {Lu}}, \bibinfo {author}
  {\bibfnamefont {S.}~\bibnamefont {Khrapak}},\ and\ \bibinfo {author}
  {\bibfnamefont {Y.}~\bibnamefont {Feng}},\ }\bibfield  {title} {\bibinfo
  {title} {Universal scaling of transverse sound speed and its isomorphic
  property in {Y}ukawa fluids},\ }\href
  {https://doi.org/10.1103/physreve.109.035202} {\bibfield  {journal} {\bibinfo
   {journal} {Phys. Rev. E}\ }\textbf {\bibinfo {volume} {109}},\ \bibinfo
  {pages} {035202} (\bibinfo {year} {2024})}\BibitemShut {NoStop}%
\bibitem [{\citenamefont {Xu}\ \emph {et~al.}(2026)\citenamefont {Xu},
  \citenamefont {Yu}, \citenamefont {Huang},\ and\ \citenamefont
  {Feng}}]{XuPRR2026}%
  \BibitemOpen
  \bibfield  {author} {\bibinfo {author} {\bibfnamefont {A.}~\bibnamefont
  {Xu}}, \bibinfo {author} {\bibfnamefont {N.}~\bibnamefont {Yu}}, \bibinfo
  {author} {\bibfnamefont {D.}~\bibnamefont {Huang}},\ and\ \bibinfo {author}
  {\bibfnamefont {Y.}~\bibnamefont {Feng}},\ }\bibfield  {title} {\bibinfo
  {title} {Identifying liquidlike and gaslike states of dusty plasmas using
  isomorph theory},\ }\bibfield  {journal} {\bibinfo  {journal} {Phys. Rev.
  Res.}\ }\textbf {\bibinfo {volume} {8}},\ \href
  {https://doi.org/10.1103/nygm-3tsj} {10.1103/nygm-3tsj} (\bibinfo {year}
  {2026})\BibitemShut {NoStop}%
\bibitem [{\citenamefont {Khrapak}(ress)}]{KhrapakPoF2026}%
  \BibitemOpen
  \bibfield  {author} {\bibinfo {author} {\bibfnamefont {S.}~\bibnamefont
  {Khrapak}},\ }\bibfield  {title} {\bibinfo {title} {Excess entropy scaling of
  the transverse sound speed in simple fluids},\ }\href@noop {} {\bibfield
  {journal} {\bibinfo  {journal} {Phys. Fluids}\ }\textbf {\bibinfo {volume}
  {8}} (\bibinfo {year} {2026, in press})}\BibitemShut {NoStop}%
\bibitem [{\citenamefont {Heyes}\ and\ \citenamefont
  {Aston}(1994)}]{HeyesJCP1994}%
  \BibitemOpen
  \bibfield  {author} {\bibinfo {author} {\bibfnamefont {D.~M.}\ \bibnamefont
  {Heyes}}\ and\ \bibinfo {author} {\bibfnamefont {P.~J.}\ \bibnamefont
  {Aston}},\ }\bibfield  {title} {\bibinfo {title} {Elastic moduli of simple
  fluids with steeply repulsive potentials},\ }\href
  {https://doi.org/10.1063/1.466511} {\bibfield  {journal} {\bibinfo  {journal}
  {J. Chem. Phys.}\ }\textbf {\bibinfo {volume} {100}},\ \bibinfo {pages}
  {2149–2153} (\bibinfo {year} {1994})}\BibitemShut {NoStop}%
\bibitem [{\citenamefont {Bamdad}\ \emph {et~al.}(2006)\citenamefont {Bamdad},
  \citenamefont {Alavi}, \citenamefont {Najafi},\ and\ \citenamefont
  {Keshavarzi}}]{BamdadChemPhys2006}%
  \BibitemOpen
  \bibfield  {author} {\bibinfo {author} {\bibfnamefont {M.}~\bibnamefont
  {Bamdad}}, \bibinfo {author} {\bibfnamefont {S.}~\bibnamefont {Alavi}},
  \bibinfo {author} {\bibfnamefont {B.}~\bibnamefont {Najafi}},\ and\ \bibinfo
  {author} {\bibfnamefont {E.}~\bibnamefont {Keshavarzi}},\ }\bibfield  {title}
  {\bibinfo {title} {A new expression for radial distribution function and
  infinite shear modulus of {L}ennard-{J}ones fluids},\ }\href
  {https://doi.org/10.1016/j.chemphys.2006.02.001} {\bibfield  {journal}
  {\bibinfo  {journal} {Chem. Phys.}\ }\textbf {\bibinfo {volume} {325}},\
  \bibinfo {pages} {554–562} (\bibinfo {year} {2006})}\BibitemShut {NoStop}%
\bibitem [{\citenamefont {Heyes}\ \emph {et~al.}(2008)\citenamefont {Heyes},
  \citenamefont {Rickayzen},\ and\ \citenamefont {Powles}}]{HeyesJCP2008}%
  \BibitemOpen
  \bibfield  {author} {\bibinfo {author} {\bibfnamefont {D.~M.}\ \bibnamefont
  {Heyes}}, \bibinfo {author} {\bibfnamefont {G.}~\bibnamefont {Rickayzen}},\
  and\ \bibinfo {author} {\bibfnamefont {J.~G.}\ \bibnamefont {Powles}},\
  }\bibfield  {title} {\bibinfo {title} {Monte {C}arlo simulations of fluids
  whose particles interact with a logarithmic potential},\ }\href
  {https://doi.org/10.1063/1.2884691} {\bibfield  {journal} {\bibinfo
  {journal} {J. Chem. Phys.}\ }\textbf {\bibinfo {volume} {128}},\ \bibinfo
  {pages} {134503} (\bibinfo {year} {2008})}\BibitemShut {NoStop}%
\bibitem [{\citenamefont {Khrapak}\ and\ \citenamefont
  {Klumov}(2020)}]{KhrapakPoP02_2020}%
  \BibitemOpen
  \bibfield  {author} {\bibinfo {author} {\bibfnamefont {S.~A.}\ \bibnamefont
  {Khrapak}}\ and\ \bibinfo {author} {\bibfnamefont {B.~A.}\ \bibnamefont
  {Klumov}},\ }\bibfield  {title} {\bibinfo {title} {Instantaneous shear
  modulus of {Y}ukawa fluids across coupling regimes},\ }\href
  {https://doi.org/10.1063/1.5140858} {\bibfield  {journal} {\bibinfo
  {journal} {Phys. Plasmas}\ }\textbf {\bibinfo {volume} {27}},\ \bibinfo
  {pages} {024501} (\bibinfo {year} {2020})}\BibitemShut {NoStop}%
\bibitem [{\citenamefont {Rogers}\ and\ \citenamefont
  {Young}(1984)}]{RogersPRA1984}%
  \BibitemOpen
  \bibfield  {author} {\bibinfo {author} {\bibfnamefont {F.~J.}\ \bibnamefont
  {Rogers}}\ and\ \bibinfo {author} {\bibfnamefont {D.~A.}\ \bibnamefont
  {Young}},\ }\bibfield  {title} {\bibinfo {title} {New, thermodynamically
  consistent, integral equation for simple fluids},\ }\href
  {https://doi.org/10.1103/physreva.30.999} {\bibfield  {journal} {\bibinfo
  {journal} {Phys. Rev. A}\ }\textbf {\bibinfo {volume} {30}},\ \bibinfo
  {pages} {999–1007} (\bibinfo {year} {1984})}\BibitemShut {NoStop}%
\bibitem [{\citenamefont {Kalman}\ \emph {et~al.}(2000)\citenamefont {Kalman},
  \citenamefont {Rosenberg},\ and\ \citenamefont {DeWitt}}]{KalmanPRL2000}%
  \BibitemOpen
  \bibfield  {author} {\bibinfo {author} {\bibfnamefont {G.}~\bibnamefont
  {Kalman}}, \bibinfo {author} {\bibfnamefont {M.}~\bibnamefont {Rosenberg}},\
  and\ \bibinfo {author} {\bibfnamefont {H.~E.}\ \bibnamefont {DeWitt}},\
  }\bibfield  {title} {\bibinfo {title} {Collective modes in strongly
  correlated {Y}ukawa liquids: {W}aves in dusty plasmas},\ }\href
  {https://doi.org/10.1103/physrevlett.84.6030} {\bibfield  {journal} {\bibinfo
   {journal} {Phys. Rev. Lett.}\ }\textbf {\bibinfo {volume} {84}},\ \bibinfo
  {pages} {6030} (\bibinfo {year} {2000})}\BibitemShut {NoStop}%
\bibitem [{\citenamefont {Tolias}\ and\ \citenamefont
  {Lucco~Castello}(2021)}]{ToliasPoP2021}%
  \BibitemOpen
  \bibfield  {author} {\bibinfo {author} {\bibfnamefont {P.}~\bibnamefont
  {Tolias}}\ and\ \bibinfo {author} {\bibfnamefont {F.}~\bibnamefont
  {Lucco~Castello}},\ }\bibfield  {title} {\bibinfo {title} {Description of
  longitudinal modes in moderately coupled {Y}ukawa systems with the static
  local field correction},\ }\href {https://doi.org/10.1063/5.0044871}
  {\bibfield  {journal} {\bibinfo  {journal} {Phys. Plasmas}\ }\textbf
  {\bibinfo {volume} {28}},\ \bibinfo {pages} {034502} (\bibinfo {year}
  {2021})}\BibitemShut {NoStop}%
\bibitem [{\citenamefont {Meier}(2002)}]{Meier2002}%
  \BibitemOpen
  \bibfield  {author} {\bibinfo {author} {\bibfnamefont {K.}~\bibnamefont
  {Meier}},\ }\href@noop {} {\emph {\bibinfo {title} {Computer Simulation and
  Interpretation of the Transport Coefficients of the {L}ennard-{J}ones Model
  Fluid (PhD Thesis)}}}\ (\bibinfo  {publisher} {Shaker},\ \bibinfo {address}
  {Aachen},\ \bibinfo {year} {2002})\BibitemShut {NoStop}%
\bibitem [{\citenamefont {Khrapak}(2020)}]{KhrapakMolecules2020}%
  \BibitemOpen
  \bibfield  {author} {\bibinfo {author} {\bibfnamefont {S.~A.}\ \bibnamefont
  {Khrapak}},\ }\bibfield  {title} {\bibinfo {title} {Sound velocities of
  {L}ennard-{J}ones systems near the liquid-solid phase transition},\ }\href
  {https://doi.org/10.3390/molecules25153498} {\bibfield  {journal} {\bibinfo
  {journal} {Molecules}\ }\textbf {\bibinfo {volume} {25}},\ \bibinfo {pages}
  {3498} (\bibinfo {year} {2020})}\BibitemShut {NoStop}%
\bibitem [{\citenamefont {Khrapak}(2021{\natexlab{d}})}]{KhrapakMolecules2021}%
  \BibitemOpen
  \bibfield  {author} {\bibinfo {author} {\bibfnamefont {S.}~\bibnamefont
  {Khrapak}},\ }\bibfield  {title} {\bibinfo {title} {Sound velocities of
  generalized {L}ennard-{J}ones (n - 6) fluids near freezing},\ }\href
  {https://doi.org/10.3390/molecules26061660} {\bibfield  {journal} {\bibinfo
  {journal} {Molecules}\ }\textbf {\bibinfo {volume} {26}},\ \bibinfo {pages}
  {1660} (\bibinfo {year} {2021}{\natexlab{d}})}\BibitemShut {NoStop}%
\bibitem [{\citenamefont {Golden}\ and\ \citenamefont
  {Kalman}(1993)}]{GoldenPSS1993}%
  \BibitemOpen
  \bibfield  {author} {\bibinfo {author} {\bibfnamefont {K.~I.}\ \bibnamefont
  {Golden}}\ and\ \bibinfo {author} {\bibfnamefont {G.}~\bibnamefont
  {Kalman}},\ }\bibfield  {title} {\bibinfo {title} {Correlations destroy
  acoustic plasmons in superlattices},\ }\href
  {https://doi.org/10.1002/pssb.2221800224} {\bibfield  {journal} {\bibinfo
  {journal} {Phys. Status Solidi (b)}\ }\textbf {\bibinfo {volume} {180}},\
  \bibinfo {pages} {533} (\bibinfo {year} {1993})}\BibitemShut {NoStop}%
\bibitem [{\citenamefont {Khrapak}\ \emph {et~al.}(2016)\citenamefont
  {Khrapak}, \citenamefont {Klumov}, \citenamefont {Couedel},\ and\
  \citenamefont {Thomas}}]{KhrapakPoP2016}%
  \BibitemOpen
  \bibfield  {author} {\bibinfo {author} {\bibfnamefont {S.~A.}\ \bibnamefont
  {Khrapak}}, \bibinfo {author} {\bibfnamefont {B.}~\bibnamefont {Klumov}},
  \bibinfo {author} {\bibfnamefont {L.}~\bibnamefont {Couedel}},\ and\ \bibinfo
  {author} {\bibfnamefont {H.~M.}\ \bibnamefont {Thomas}},\ }\bibfield  {title}
  {\bibinfo {title} {On the long-waves dispersion in {Y}ukawa systems},\ }\href
  {https://doi.org/10.1063/1.4942169} {\bibfield  {journal} {\bibinfo
  {journal} {Phys. Plasmas}\ }\textbf {\bibinfo {volume} {23}},\ \bibinfo
  {pages} {023702} (\bibinfo {year} {2016})}\BibitemShut {NoStop}%
\bibitem [{\citenamefont {Khrapak}(2017)}]{KhrapakAIPAdv2017}%
  \BibitemOpen
  \bibfield  {author} {\bibinfo {author} {\bibfnamefont {S.~A.}\ \bibnamefont
  {Khrapak}},\ }\bibfield  {title} {\bibinfo {title} {Practical dispersion
  relations for strongly coupled plasma fluids},\ }\href
  {https://doi.org/10.1063/1.5002130} {\bibfield  {journal} {\bibinfo
  {journal} {{AIP} Adv.}\ }\textbf {\bibinfo {volume} {7}},\ \bibinfo {pages}
  {125026} (\bibinfo {year} {2017})}\BibitemShut {NoStop}%
\bibitem [{\citenamefont {Khrapak}\ and\ \citenamefont
  {Khrapak}(2018)}]{KhrapakIEEE2018}%
  \BibitemOpen
  \bibfield  {author} {\bibinfo {author} {\bibfnamefont {S.}~\bibnamefont
  {Khrapak}}\ and\ \bibinfo {author} {\bibfnamefont {A.}~\bibnamefont
  {Khrapak}},\ }\bibfield  {title} {\bibinfo {title} {Simple dispersion
  relations for {C}oulomb and {Y}ukawa fluids},\ }\href
  {https://doi.org/10.1109/tps.2017.2763741} {\bibfield  {journal} {\bibinfo
  {journal} {{IEEE} Trans. Plasma Sci.}\ }\textbf {\bibinfo {volume} {46}},\
  \bibinfo {pages} {737} (\bibinfo {year} {2018})}\BibitemShut {NoStop}%
\bibitem [{\citenamefont {Fairushin}\ \emph {et~al.}(2020)\citenamefont
  {Fairushin}, \citenamefont {Khrapak},\ and\ \citenamefont
  {Mokshin}}]{FairushinResPhys2020}%
  \BibitemOpen
  \bibfield  {author} {\bibinfo {author} {\bibfnamefont {I.}~\bibnamefont
  {Fairushin}}, \bibinfo {author} {\bibfnamefont {S.}~\bibnamefont {Khrapak}},\
  and\ \bibinfo {author} {\bibfnamefont {A.}~\bibnamefont {Mokshin}},\
  }\bibfield  {title} {\bibinfo {title} {Direct evaluation of the physical
  characteristics of {Y}ukawa fluids based on a simple approximation for the
  radial distribution function},\ }\href
  {https://doi.org/10.1016/j.rinp.2020.103359} {\bibfield  {journal} {\bibinfo
  {journal} {Results Phys.}\ }\textbf {\bibinfo {volume} {19}},\ \bibinfo
  {pages} {103359} (\bibinfo {year} {2020})}\BibitemShut {NoStop}%
\bibitem [{\citenamefont {Fairushin}\ and\ \citenamefont
  {Mokshin}(2023)}]{FairushinFluids2023}%
  \BibitemOpen
  \bibfield  {author} {\bibinfo {author} {\bibfnamefont {I.~I.}\ \bibnamefont
  {Fairushin}}\ and\ \bibinfo {author} {\bibfnamefont {A.~V.}\ \bibnamefont
  {Mokshin}},\ }\bibfield  {title} {\bibinfo {title} {Calculation of
  thermodynamic characteristics and sound velocity for two-dimensional {Y}ukawa
  fluids based on a two-step approximation for the radial distribution
  function},\ }\href {https://doi.org/10.3390/fluids8020072} {\bibfield
  {journal} {\bibinfo  {journal} {Fluids}\ }\textbf {\bibinfo {volume} {8}},\
  \bibinfo {pages} {72} (\bibinfo {year} {2023})}\BibitemShut {NoStop}%
\bibitem [{\citenamefont {{Tsytovich}}(1997)}]{TsytovichUFN1997}%
  \BibitemOpen
  \bibfield  {author} {\bibinfo {author} {\bibfnamefont {V.}~\bibnamefont
  {{Tsytovich}}},\ }\bibfield  {title} {\bibinfo {title} {{Dust plasma
  crystals, drops, and clouds.}},\ }\href
  {https://doi.org/10.1070/PU1997v040n01ABEH000201} {\bibfield  {journal}
  {\bibinfo  {journal} {Phys.-Usp.}\ }\textbf {\bibinfo {volume} {40}},\
  \bibinfo {pages} {53} (\bibinfo {year} {1997})}\BibitemShut {NoStop}%
\bibitem [{\citenamefont {Shukla}\ and\ \citenamefont
  {Eliasson}(2009)}]{ShuklaRMP2009}%
  \BibitemOpen
  \bibfield  {author} {\bibinfo {author} {\bibfnamefont {P.~K.}\ \bibnamefont
  {Shukla}}\ and\ \bibinfo {author} {\bibfnamefont {B.}~\bibnamefont
  {Eliasson}},\ }\bibfield  {title} {\bibinfo {title} {Colloquium: Fundamentals
  of dust-plasma interactions},\ }\href
  {https://doi.org/10.1103/revmodphys.81.25} {\bibfield  {journal} {\bibinfo
  {journal} {Rev. Mod. Phys.}\ }\textbf {\bibinfo {volume} {81}},\ \bibinfo
  {pages} {25–44} (\bibinfo {year} {2009})}\BibitemShut {NoStop}%
\bibitem [{\citenamefont {Fortov}\ \emph {et~al.}(2005)\citenamefont {Fortov},
  \citenamefont {Ivlev}, \citenamefont {Khrapak}, \citenamefont {Khrapak},\
  and\ \citenamefont {Morfill}}]{FortovPR}%
  \BibitemOpen
  \bibfield  {author} {\bibinfo {author} {\bibfnamefont {V.~E.}\ \bibnamefont
  {Fortov}}, \bibinfo {author} {\bibfnamefont {A.~V.}\ \bibnamefont {Ivlev}},
  \bibinfo {author} {\bibfnamefont {S.~A.}\ \bibnamefont {Khrapak}}, \bibinfo
  {author} {\bibfnamefont {A.~G.}\ \bibnamefont {Khrapak}},\ and\ \bibinfo
  {author} {\bibfnamefont {G.~E.}\ \bibnamefont {Morfill}},\ }\bibfield
  {title} {\bibinfo {title} {Complex (dusty) plasmas: Current status, open
  issues, perspectives},\ }\href@noop {} {\bibfield  {journal} {\bibinfo
  {journal} {Phys. Rep.}\ }\textbf {\bibinfo {volume} {421}},\ \bibinfo {pages}
  {1} (\bibinfo {year} {2005})}\BibitemShut {NoStop}%
\bibitem [{\citenamefont {Morfill}\ and\ \citenamefont
  {Ivlev}(2009)}]{MorfillRMP2009}%
  \BibitemOpen
  \bibfield  {author} {\bibinfo {author} {\bibfnamefont {G.~E.}\ \bibnamefont
  {Morfill}}\ and\ \bibinfo {author} {\bibfnamefont {A.~V.}\ \bibnamefont
  {Ivlev}},\ }\bibfield  {title} {\bibinfo {title} {Complex plasmas: An
  interdisciplinary research field},\ }\href
  {https://doi.org/10.1103/revmodphys.81.1353} {\bibfield  {journal} {\bibinfo
  {journal} {Rev. Mod. Phys.}\ }\textbf {\bibinfo {volume} {81}},\ \bibinfo
  {pages} {1353–1404} (\bibinfo {year} {2009})}\BibitemShut {NoStop}%
\bibitem [{\citenamefont {Bonitz}\ \emph {et~al.}(2010)\citenamefont {Bonitz},
  \citenamefont {Henning},\ and\ \citenamefont {Block}}]{BonitzRPP2010}%
  \BibitemOpen
  \bibfield  {author} {\bibinfo {author} {\bibfnamefont {M.}~\bibnamefont
  {Bonitz}}, \bibinfo {author} {\bibfnamefont {C.}~\bibnamefont {Henning}},\
  and\ \bibinfo {author} {\bibfnamefont {D.}~\bibnamefont {Block}},\ }\bibfield
   {title} {\bibinfo {title} {Complex plasmas: a laboratory for strong
  correlations},\ }\href {https://doi.org/10.1088/0034-4885/73/6/066501}
  {\bibfield  {journal} {\bibinfo  {journal} {Rep. Progr. Phys.}\ }\textbf
  {\bibinfo {volume} {73}},\ \bibinfo {pages} {066501} (\bibinfo {year}
  {2010})}\BibitemShut {NoStop}%
\bibitem [{\citenamefont {Fortov}\ and\ \citenamefont
  {Morfill}(2010)}]{FortovBook}%
  \BibitemOpen
  \bibfield  {author} {\bibinfo {author} {\bibfnamefont {V.~E.}\ \bibnamefont
  {Fortov}}\ and\ \bibinfo {author} {\bibfnamefont {G.~E.}\ \bibnamefont
  {Morfill}},\ }\href@noop {} {\emph {\bibinfo {title} {Complex and Dusty
  Plasmas - From Laboratory to Space}}}\ (\bibinfo  {publisher} {CRC Press
  LLC},\ \bibinfo {address} {Boca Raton},\ \bibinfo {year} {2010})\BibitemShut
  {NoStop}%
\bibitem [{\citenamefont {Ivlev}\ \emph {et~al.}(2012)\citenamefont {Ivlev},
  \citenamefont {L\"owen}, \citenamefont {Morfill},\ and\ \citenamefont
  {Royall}}]{IvlevBook}%
  \BibitemOpen
  \bibfield  {author} {\bibinfo {author} {\bibfnamefont {A.}~\bibnamefont
  {Ivlev}}, \bibinfo {author} {\bibfnamefont {H.}~\bibnamefont {L\"owen}},
  \bibinfo {author} {\bibfnamefont {G.}~\bibnamefont {Morfill}},\ and\ \bibinfo
  {author} {\bibfnamefont {C.~P.}\ \bibnamefont {Royall}},\ }\href@noop {}
  {\emph {\bibinfo {title} {Complex Plasmas and Colloidal Dispersions:
  Particle-Resolved Studies of Classical Liquids and Solids}}}\ (\bibinfo
  {publisher} {World Scientific},\ \bibinfo {year} {2012})\BibitemShut
  {NoStop}%
\bibitem [{\citenamefont {Melzer}(2019)}]{MelzerBook}%
  \BibitemOpen
  \bibfield  {author} {\bibinfo {author} {\bibfnamefont {A.}~\bibnamefont
  {Melzer}},\ }\href {https://doi.org/10.1007/978-3-030-20260-6} {\emph
  {\bibinfo {title} {Physics of Dusty Plasmas: {A}n Introduction}}}\ (\bibinfo
  {publisher} {Springer International Publishing},\ \bibinfo {year}
  {2019})\BibitemShut {NoStop}%
\bibitem [{\citenamefont {DeWitt}\ and\ \citenamefont
  {Rosenfeld}(1979)}]{DeWittPLA1979}%
  \BibitemOpen
  \bibfield  {author} {\bibinfo {author} {\bibfnamefont {H.}~\bibnamefont
  {DeWitt}}\ and\ \bibinfo {author} {\bibfnamefont {Y.}~\bibnamefont
  {Rosenfeld}},\ }\bibfield  {title} {\bibinfo {title} {Derivation of the one
  component plasma fluid equation of state in strong coupling},\ }\href
  {https://doi.org/10.1016/0375-9601(79)90283-4} {\bibfield  {journal}
  {\bibinfo  {journal} {Phys. Lett. A}\ }\textbf {\bibinfo {volume} {75}},\
  \bibinfo {pages} {79–80} (\bibinfo {year} {1979})}\BibitemShut {NoStop}%
\bibitem [{\citenamefont {Khrapak}\ and\ \citenamefont
  {Khrapak}(2026)}]{YukawaSubmitted}%
  \BibitemOpen
  \bibfield  {author} {\bibinfo {author} {\bibfnamefont {S.~A.}\ \bibnamefont
  {Khrapak}}\ and\ \bibinfo {author} {\bibfnamefont {A.~G.}\ \bibnamefont
  {Khrapak}},\ }\bibfield  {title} {\bibinfo {title} {Variational approach to
  {Y}ukawa fluids. {I}. {T}hermodynamics},\ }\href
  {https://doi.org/10.1103/kvnp-nf1l} {\bibfield  {journal} {\bibinfo
  {journal} {Phys. Rev. E}\ }\textbf {\bibinfo {volume} {114}},\ \bibinfo
  {pages} {015216} (\bibinfo {year} {2026})}\BibitemShut {NoStop}%
\bibitem [{\citenamefont {Wertheim}(1963)}]{Wertheim1963}%
  \BibitemOpen
  \bibfield  {author} {\bibinfo {author} {\bibfnamefont {M.~S.}\ \bibnamefont
  {Wertheim}},\ }\bibfield  {title} {\bibinfo {title} {Exact solution of the
  {Percus-Yevick} integral equation for hard spheres},\ }\href
  {https://doi.org/10.1103/physrevlett.10.321} {\bibfield  {journal} {\bibinfo
  {journal} {Phys. Rev. Lett.}\ }\textbf {\bibinfo {volume} {10}},\ \bibinfo
  {pages} {321} (\bibinfo {year} {1963})}\BibitemShut {NoStop}%
\bibitem [{\citenamefont {Thiele}(1963)}]{Thiele1963}%
  \BibitemOpen
  \bibfield  {author} {\bibinfo {author} {\bibfnamefont {E.}~\bibnamefont
  {Thiele}},\ }\bibfield  {title} {\bibinfo {title} {Equation of state for hard
  spheres},\ }\href {https://doi.org/10.1063/1.1734272} {\bibfield  {journal}
  {\bibinfo  {journal} {J. Chem. Phys.}\ }\textbf {\bibinfo {volume} {39}},\
  \bibinfo {pages} {474–479} (\bibinfo {year} {1963})}\BibitemShut {NoStop}%
\bibitem [{\citenamefont {Khrapak}\ and\ \citenamefont
  {Khrapak}(2023{\natexlab{b}})}]{KhrapakPoF2023}%
  \BibitemOpen
  \bibfield  {author} {\bibinfo {author} {\bibfnamefont {S.~A.}\ \bibnamefont
  {Khrapak}}\ and\ \bibinfo {author} {\bibfnamefont {A.~G.}\ \bibnamefont
  {Khrapak}},\ }\bibfield  {title} {\bibinfo {title} {Sound velocities in
  liquids near freezing: {D}ependence on the interaction potential and
  correlations with thermal conductivity},\ }\href
  {https://doi.org/10.1063/5.0157945} {\bibfield  {journal} {\bibinfo
  {journal} {Phys. Fluids}\ }\textbf {\bibinfo {volume} {35}},\ \bibinfo
  {pages} {077129} (\bibinfo {year} {2023}{\natexlab{b}})}\BibitemShut
  {NoStop}%
\bibitem [{\citenamefont {Donko}\ \emph {et~al.}(2008)\citenamefont {Donko},
  \citenamefont {Kalman},\ and\ \citenamefont {Hartmann}}]{DonkoJPCM2008}%
  \BibitemOpen
  \bibfield  {author} {\bibinfo {author} {\bibfnamefont {Z.}~\bibnamefont
  {Donko}}, \bibinfo {author} {\bibfnamefont {G.~J.}\ \bibnamefont {Kalman}},\
  and\ \bibinfo {author} {\bibfnamefont {P.}~\bibnamefont {Hartmann}},\
  }\bibfield  {title} {\bibinfo {title} {Dynamical correlations and collective
  excitations of {Y}ukawa liquids},\ }\href
  {https://doi.org/10.1088/0953-8984/20/41/413101} {\bibfield  {journal}
  {\bibinfo  {journal} {J. Phys.: Condens. Matter}\ }\textbf {\bibinfo {volume}
  {20}},\ \bibinfo {pages} {413101} (\bibinfo {year} {2008})}\BibitemShut
  {NoStop}%
\bibitem [{\citenamefont {Khrapak}(2016)}]{KhrapakPoP02_2016_1}%
  \BibitemOpen
  \bibfield  {author} {\bibinfo {author} {\bibfnamefont {S.~A.}\ \bibnamefont
  {Khrapak}},\ }\bibfield  {title} {\bibinfo {title} {Relations between the
  longitudinal and transverse sound velocities in strongly coupled {Y}ukawa
  fluids},\ }\href {https://doi.org/10.1063/1.4942171} {\bibfield  {journal}
  {\bibinfo  {journal} {Phys. Plasmas}\ }\textbf {\bibinfo {volume} {23}},\
  \bibinfo {pages} {024504} (\bibinfo {year} {2016})}\BibitemShut {NoStop}%
\bibitem [{\citenamefont {Brazhkin}\ \emph
  {et~al.}(2012{\natexlab{a}})\citenamefont {Brazhkin}, \citenamefont {Fomin},
  \citenamefont {Lyapin}, \citenamefont {Ryzhov},\ and\ \citenamefont
  {Trachenko}}]{BrazhkinPRE2012}%
  \BibitemOpen
  \bibfield  {author} {\bibinfo {author} {\bibfnamefont {V.~V.}\ \bibnamefont
  {Brazhkin}}, \bibinfo {author} {\bibfnamefont {Y.~D.}\ \bibnamefont {Fomin}},
  \bibinfo {author} {\bibfnamefont {A.~G.}\ \bibnamefont {Lyapin}}, \bibinfo
  {author} {\bibfnamefont {V.~N.}\ \bibnamefont {Ryzhov}},\ and\ \bibinfo
  {author} {\bibfnamefont {K.}~\bibnamefont {Trachenko}},\ }\bibfield  {title}
  {\bibinfo {title} {Two liquid states of matter: A dynamic line on a phase
  diagram},\ }\href {https://doi.org/10.1103/physreve.85.031203} {\bibfield
  {journal} {\bibinfo  {journal} {Phys. Rev. E}\ }\textbf {\bibinfo {volume}
  {85}},\ \bibinfo {pages} {031203} (\bibinfo {year}
  {2012}{\natexlab{a}})}\BibitemShut {NoStop}%
\bibitem [{\citenamefont {Brazhkin}\ \emph
  {et~al.}(2012{\natexlab{b}})\citenamefont {Brazhkin}, \citenamefont {Lyapin},
  \citenamefont {Ryzhov}, \citenamefont {Trachenko}, \citenamefont {Fomin},\
  and\ \citenamefont {Tsiok}}]{BrazhkinUFN2012}%
  \BibitemOpen
  \bibfield  {author} {\bibinfo {author} {\bibfnamefont {V.~V.}\ \bibnamefont
  {Brazhkin}}, \bibinfo {author} {\bibfnamefont {A.}~\bibnamefont {Lyapin}},
  \bibinfo {author} {\bibfnamefont {V.~N.}\ \bibnamefont {Ryzhov}}, \bibinfo
  {author} {\bibfnamefont {K.}~\bibnamefont {Trachenko}}, \bibinfo {author}
  {\bibfnamefont {Y.~D.}\ \bibnamefont {Fomin}},\ and\ \bibinfo {author}
  {\bibfnamefont {E.~N.}\ \bibnamefont {Tsiok}},\ }\bibfield  {title} {\bibinfo
  {title} {Where is the supercritical fluid on the phase diagram?},\ }\href
  {https://doi.org/10.3367/ufnr.0182.201211a.1137} {\bibfield  {journal}
  {\bibinfo  {journal} {Phys.-Usp.}\ }\textbf {\bibinfo {volume} {182}},\
  \bibinfo {pages} {1137} (\bibinfo {year} {2012}{\natexlab{b}})}\BibitemShut
  {NoStop}%
\bibitem [{\citenamefont {Brazhkin}\ \emph {et~al.}(2013)\citenamefont
  {Brazhkin}, \citenamefont {Fomin}, \citenamefont {Lyapin}, \citenamefont
  {Ryzhov}, \citenamefont {Tsiok},\ and\ \citenamefont
  {Trachenko}}]{BrazhkinPRL2013}%
  \BibitemOpen
  \bibfield  {author} {\bibinfo {author} {\bibfnamefont {V.~V.}\ \bibnamefont
  {Brazhkin}}, \bibinfo {author} {\bibfnamefont {Y.~D.}\ \bibnamefont {Fomin}},
  \bibinfo {author} {\bibfnamefont {A.~G.}\ \bibnamefont {Lyapin}}, \bibinfo
  {author} {\bibfnamefont {V.~N.}\ \bibnamefont {Ryzhov}}, \bibinfo {author}
  {\bibfnamefont {E.~N.}\ \bibnamefont {Tsiok}},\ and\ \bibinfo {author}
  {\bibfnamefont {K.}~\bibnamefont {Trachenko}},\ }\bibfield  {title} {\bibinfo
  {title} {Liquid-gas'' transition in the supercritical region: fundamental
  changes in the particle dynamics},\ }\href@noop {} {\bibfield  {journal}
  {\bibinfo  {journal} {Phys. Rev. Lett.}\ }\textbf {\bibinfo {volume} {111}},\
  \bibinfo {pages} {145901} (\bibinfo {year} {2013})}\BibitemShut {NoStop}%
\bibitem [{\citenamefont {Khrapak}(2019{\natexlab{b}})}]{KhrapakPoP10_2019}%
  \BibitemOpen
  \bibfield  {author} {\bibinfo {author} {\bibfnamefont {S.~A.}\ \bibnamefont
  {Khrapak}},\ }\bibfield  {title} {\bibinfo {title} {Unified description of
  sound velocities in strongly coupled {Y}ukawa systems of different spatial
  dimensionality},\ }\href {https://doi.org/10.1063/1.5124676} {\bibfield
  {journal} {\bibinfo  {journal} {Phys. Plasmas}\ }\textbf {\bibinfo {volume}
  {26}},\ \bibinfo {pages} {103703} (\bibinfo {year}
  {2019}{\natexlab{b}})}\BibitemShut {NoStop}%
\bibitem [{\citenamefont {Rosenfeld}(2000)}]{RosenfeldPRE2000}%
  \BibitemOpen
  \bibfield  {author} {\bibinfo {author} {\bibfnamefont {Y.}~\bibnamefont
  {Rosenfeld}},\ }\bibfield  {title} {\bibinfo {title} {Excess-entropy and
  freezing-temperature scalings for transport coefficients: Self-diffusion in
  {Y}ukawa systems},\ }\href {https://doi.org/10.1103/physreve.62.7524}
  {\bibfield  {journal} {\bibinfo  {journal} {Phys. Rev. E}\ }\textbf {\bibinfo
  {volume} {62}},\ \bibinfo {pages} {7524} (\bibinfo {year}
  {2000})}\BibitemShut {NoStop}%
\bibitem [{\citenamefont {Rao}\ \emph {et~al.}(1990)\citenamefont {Rao},
  \citenamefont {Shukla},\ and\ \citenamefont {Yu}}]{Rao1990}%
  \BibitemOpen
  \bibfield  {author} {\bibinfo {author} {\bibfnamefont {N.}~\bibnamefont
  {Rao}}, \bibinfo {author} {\bibfnamefont {P.}~\bibnamefont {Shukla}},\ and\
  \bibinfo {author} {\bibfnamefont {M.}~\bibnamefont {Yu}},\ }\bibfield
  {title} {\bibinfo {title} {Dust-acoustic waves in dusty plasmas},\ }\href
  {https://doi.org/10.1016/0032-0633(90)90147-i} {\bibfield  {journal}
  {\bibinfo  {journal} {Planet. Space Sci.}\ }\textbf {\bibinfo {volume}
  {38}},\ \bibinfo {pages} {543} (\bibinfo {year} {1990})}\BibitemShut
  {NoStop}%
\bibitem [{\citenamefont {Landau}\ and\ \citenamefont
  {Lifshitz}(1987)}]{LL_Fluids}%
  \BibitemOpen
  \bibfield  {author} {\bibinfo {author} {\bibfnamefont {L.~D.}\ \bibnamefont
  {Landau}}\ and\ \bibinfo {author} {\bibfnamefont {E.~M.}\ \bibnamefont
  {Lifshitz}},\ }\href@noop {} {\emph {\bibinfo {title} {Fluid Mechanics}}}\
  (\bibinfo  {publisher} {Butterworth-Heinemann, Oxford},\ \bibinfo {year}
  {1987})\BibitemShut {NoStop}%
\bibitem [{\citenamefont {Khrapak}(2025)}]{KhrapakPRE2025}%
  \BibitemOpen
  \bibfield  {author} {\bibinfo {author} {\bibfnamefont {S.~A.}\ \bibnamefont
  {Khrapak}},\ }\bibfield  {title} {\bibinfo {title} {Speed of sound in dense
  simple liquids},\ }\href {https://doi.org/10.1103/5dtk-4x7m} {\bibfield
  {journal} {\bibinfo  {journal} {Phys. Rev. E}\ }\textbf {\bibinfo {volume}
  {111}},\ \bibinfo {pages} {065423} (\bibinfo {year} {2025})}\BibitemShut
  {NoStop}%
\bibitem [{\citenamefont {Khrapak}\ and\ \citenamefont
  {Thomas}(2015{\natexlab{a}})}]{KhrapakPRE03_2015}%
  \BibitemOpen
  \bibfield  {author} {\bibinfo {author} {\bibfnamefont {S.~A.}\ \bibnamefont
  {Khrapak}}\ and\ \bibinfo {author} {\bibfnamefont {H.~M.}\ \bibnamefont
  {Thomas}},\ }\bibfield  {title} {\bibinfo {title} {Fluid approach to evaluate
  sound velocity in {Y}ukawa systems and complex plasmas},\ }\href
  {https://doi.org/10.1103/physreve.91.033110} {\bibfield  {journal} {\bibinfo
  {journal} {Phys. Rev. E}\ }\textbf {\bibinfo {volume} {91}},\ \bibinfo
  {pages} {033110} (\bibinfo {year} {2015}{\natexlab{a}})}\BibitemShut
  {NoStop}%
\bibitem [{\citenamefont {Barkan}\ \emph {et~al.}(1995)\citenamefont {Barkan},
  \citenamefont {Merlino},\ and\ \citenamefont {D’Angelo}}]{BarkanPoP1995}%
  \BibitemOpen
  \bibfield  {author} {\bibinfo {author} {\bibfnamefont {A.}~\bibnamefont
  {Barkan}}, \bibinfo {author} {\bibfnamefont {R.~L.}\ \bibnamefont
  {Merlino}},\ and\ \bibinfo {author} {\bibfnamefont {N.}~\bibnamefont
  {D’Angelo}},\ }\bibfield  {title} {\bibinfo {title} {Laboratory observation
  of the dust-acoustic wave mode},\ }\href {https://doi.org/10.1063/1.871121}
  {\bibfield  {journal} {\bibinfo  {journal} {Phys. Plasmas}\ }\textbf
  {\bibinfo {volume} {2}},\ \bibinfo {pages} {3563–3565} (\bibinfo {year}
  {1995})}\BibitemShut {NoStop}%
\bibitem [{\citenamefont {Thompson}\ \emph {et~al.}(1997)\citenamefont
  {Thompson}, \citenamefont {Barkan}, \citenamefont {D’Angelo},\ and\
  \citenamefont {Merlino}}]{ThompsonPoP1997}%
  \BibitemOpen
  \bibfield  {author} {\bibinfo {author} {\bibfnamefont {C.}~\bibnamefont
  {Thompson}}, \bibinfo {author} {\bibfnamefont {A.}~\bibnamefont {Barkan}},
  \bibinfo {author} {\bibfnamefont {N.}~\bibnamefont {D’Angelo}},\ and\
  \bibinfo {author} {\bibfnamefont {R.~L.}\ \bibnamefont {Merlino}},\
  }\bibfield  {title} {\bibinfo {title} {Dust acoustic waves in a direct
  current glow discharge},\ }\href {https://doi.org/10.1063/1.872238}
  {\bibfield  {journal} {\bibinfo  {journal} {Phys. Plasmas}\ }\textbf
  {\bibinfo {volume} {4}},\ \bibinfo {pages} {2331–2335} (\bibinfo {year}
  {1997})}\BibitemShut {NoStop}%
\bibitem [{\citenamefont {Khrapak}\ \emph {et~al.}(2003)\citenamefont
  {Khrapak}, \citenamefont {Samsonov}, \citenamefont {Morfill}, \citenamefont
  {Thomas}, \citenamefont {Yaroshenko}, \citenamefont {Rothermel},
  \citenamefont {Hagl}, \citenamefont {Fortov}, \citenamefont {Nefedov},
  \citenamefont {Molotkov}, \citenamefont {Petrov}, \citenamefont {Lipaev},
  \citenamefont {Ivanov},\ and\ \citenamefont {Baturin}}]{KhrapakPoP01_2003}%
  \BibitemOpen
  \bibfield  {author} {\bibinfo {author} {\bibfnamefont {S.}~\bibnamefont
  {Khrapak}}, \bibinfo {author} {\bibfnamefont {D.}~\bibnamefont {Samsonov}},
  \bibinfo {author} {\bibfnamefont {G.}~\bibnamefont {Morfill}}, \bibinfo
  {author} {\bibfnamefont {H.}~\bibnamefont {Thomas}}, \bibinfo {author}
  {\bibfnamefont {V.}~\bibnamefont {Yaroshenko}}, \bibinfo {author}
  {\bibfnamefont {H.}~\bibnamefont {Rothermel}}, \bibinfo {author}
  {\bibfnamefont {T.}~\bibnamefont {Hagl}}, \bibinfo {author} {\bibfnamefont
  {V.}~\bibnamefont {Fortov}}, \bibinfo {author} {\bibfnamefont
  {A.}~\bibnamefont {Nefedov}}, \bibinfo {author} {\bibfnamefont
  {V.}~\bibnamefont {Molotkov}}, \bibinfo {author} {\bibfnamefont
  {O.}~\bibnamefont {Petrov}}, \bibinfo {author} {\bibfnamefont
  {A.}~\bibnamefont {Lipaev}}, \bibinfo {author} {\bibfnamefont
  {A.}~\bibnamefont {Ivanov}},\ and\ \bibinfo {author} {\bibfnamefont
  {Y.}~\bibnamefont {Baturin}},\ }\bibfield  {title} {\bibinfo {title}
  {Compressional waves in complex (dusty) plasmas under microgravity
  conditions},\ }\href {https://doi.org/10.1063/1.1525283} {\bibfield
  {journal} {\bibinfo  {journal} {Phys. Plasmas}\ }\textbf {\bibinfo {volume}
  {10}},\ \bibinfo {pages} {1–4} (\bibinfo {year} {2003})}\BibitemShut
  {NoStop}%
\bibitem [{\citenamefont {Annibaldi}\ \emph {et~al.}(2007)\citenamefont
  {Annibaldi}, \citenamefont {Ivlev}, \citenamefont {Konopka}, \citenamefont
  {Ratynskaia}, \citenamefont {Thomas}, \citenamefont {Morfill}, \citenamefont
  {Lipaev}, \citenamefont {Molotkov}, \citenamefont {Petrov},\ and\
  \citenamefont {Fortov}}]{AnnibaldiNJP2007}%
  \BibitemOpen
  \bibfield  {author} {\bibinfo {author} {\bibfnamefont {S.~V.}\ \bibnamefont
  {Annibaldi}}, \bibinfo {author} {\bibfnamefont {A.~V.}\ \bibnamefont
  {Ivlev}}, \bibinfo {author} {\bibfnamefont {U.}~\bibnamefont {Konopka}},
  \bibinfo {author} {\bibfnamefont {S.}~\bibnamefont {Ratynskaia}}, \bibinfo
  {author} {\bibfnamefont {H.~M.}\ \bibnamefont {Thomas}}, \bibinfo {author}
  {\bibfnamefont {G.~E.}\ \bibnamefont {Morfill}}, \bibinfo {author}
  {\bibfnamefont {A.~M.}\ \bibnamefont {Lipaev}}, \bibinfo {author}
  {\bibfnamefont {V.~I.}\ \bibnamefont {Molotkov}}, \bibinfo {author}
  {\bibfnamefont {O.~F.}\ \bibnamefont {Petrov}},\ and\ \bibinfo {author}
  {\bibfnamefont {V.~E.}\ \bibnamefont {Fortov}},\ }\bibfield  {title}
  {\bibinfo {title} {Dust-acoustic dispersion relation in three-dimensional
  complex plasmas under microgravity},\ }\href
  {https://doi.org/10.1088/1367-2630/9/9/327} {\bibfield  {journal} {\bibinfo
  {journal} {New J. Phys.}\ }\textbf {\bibinfo {volume} {9}},\ \bibinfo {pages}
  {327–327} (\bibinfo {year} {2007})}\BibitemShut {NoStop}%
\bibitem [{\citenamefont {Schwabe}\ \emph {et~al.}(2011)\citenamefont
  {Schwabe}, \citenamefont {Jiang}, \citenamefont {Zhdanov}, \citenamefont
  {Hagl}, \citenamefont {Huber}, \citenamefont {Ivlev}, \citenamefont {Lipaev},
  \citenamefont {Molotkov}, \citenamefont {Naumkin}, \citenamefont
  {S\"{u}tterlin}, \citenamefont {Thomas}, \citenamefont {Fortov},
  \citenamefont {Morfill}, \citenamefont {Skvortsov},\ and\ \citenamefont
  {Volkov}}]{SchwabeEPL2011}%
  \BibitemOpen
  \bibfield  {author} {\bibinfo {author} {\bibfnamefont {M.}~\bibnamefont
  {Schwabe}}, \bibinfo {author} {\bibfnamefont {K.}~\bibnamefont {Jiang}},
  \bibinfo {author} {\bibfnamefont {S.}~\bibnamefont {Zhdanov}}, \bibinfo
  {author} {\bibfnamefont {T.}~\bibnamefont {Hagl}}, \bibinfo {author}
  {\bibfnamefont {P.}~\bibnamefont {Huber}}, \bibinfo {author} {\bibfnamefont
  {A.~V.}\ \bibnamefont {Ivlev}}, \bibinfo {author} {\bibfnamefont {A.~M.}\
  \bibnamefont {Lipaev}}, \bibinfo {author} {\bibfnamefont {V.~I.}\
  \bibnamefont {Molotkov}}, \bibinfo {author} {\bibfnamefont {V.~N.}\
  \bibnamefont {Naumkin}}, \bibinfo {author} {\bibfnamefont {K.~R.}\
  \bibnamefont {S\"{u}tterlin}}, \bibinfo {author} {\bibfnamefont {H.~M.}\
  \bibnamefont {Thomas}}, \bibinfo {author} {\bibfnamefont {V.~E.}\
  \bibnamefont {Fortov}}, \bibinfo {author} {\bibfnamefont {G.~E.}\
  \bibnamefont {Morfill}}, \bibinfo {author} {\bibfnamefont {A.}~\bibnamefont
  {Skvortsov}},\ and\ \bibinfo {author} {\bibfnamefont {S.}~\bibnamefont
  {Volkov}},\ }\bibfield  {title} {\bibinfo {title} {Direct measurement of the
  speed of sound in a complex plasma under microgravity conditions},\ }\href
  {https://doi.org/10.1209/0295-5075/96/55001} {\bibfield  {journal} {\bibinfo
  {journal} {EPL (Europhys. Lett.)}\ }\textbf {\bibinfo {volume} {96}},\
  \bibinfo {pages} {55001} (\bibinfo {year} {2011})}\BibitemShut {NoStop}%
\bibitem [{\citenamefont {Delzanno}\ and\ \citenamefont
  {Lapenta}(2005)}]{DelzannoPRL2005}%
  \BibitemOpen
  \bibfield  {author} {\bibinfo {author} {\bibfnamefont {G.~L.}\ \bibnamefont
  {Delzanno}}\ and\ \bibinfo {author} {\bibfnamefont {G.}~\bibnamefont
  {Lapenta}},\ }\bibfield  {title} {\bibinfo {title} {Modified {J}eans
  instability for dust grains in a plasma},\ }\href
  {https://doi.org/10.1103/physrevlett.94.175005} {\bibfield  {journal}
  {\bibinfo  {journal} {Phys. Rev. Lett.}\ }\textbf {\bibinfo {volume} {94}},\
  \bibinfo {pages} {175005} (\bibinfo {year} {2005})}\BibitemShut {NoStop}%
\bibitem [{\citenamefont {Khrapak}\ \emph
  {et~al.}(2017{\natexlab{b}})\citenamefont {Khrapak}, \citenamefont {Klumov},\
  and\ \citenamefont {Thomas}}]{KhrapakPoP02_2017}%
  \BibitemOpen
  \bibfield  {author} {\bibinfo {author} {\bibfnamefont {S.~A.}\ \bibnamefont
  {Khrapak}}, \bibinfo {author} {\bibfnamefont {B.~A.}\ \bibnamefont
  {Klumov}},\ and\ \bibinfo {author} {\bibfnamefont {H.~M.}\ \bibnamefont
  {Thomas}},\ }\bibfield  {title} {\bibinfo {title} {Fingerprints of different
  interaction mechanisms on the collective modes in complex (dusty) plasmas},\
  }\href {https://doi.org/10.1063/1.4976124} {\bibfield  {journal} {\bibinfo
  {journal} {Phys. Plasmas}\ }\textbf {\bibinfo {volume} {24}},\ \bibinfo
  {pages} {023702} (\bibinfo {year} {2017}{\natexlab{b}})}\BibitemShut
  {NoStop}%
\bibitem [{\citenamefont {Schwabe}\ \emph {et~al.}(2020)\citenamefont
  {Schwabe}, \citenamefont {Khrapak}, \citenamefont {Zhdanov}, \citenamefont
  {Pustylnik}, \citenamefont {R\"{a}th}, \citenamefont {Fink}, \citenamefont
  {Kretschmer}, \citenamefont {Lipaev}, \citenamefont {Molotkov}, \citenamefont
  {Schmitz}, \citenamefont {Thoma}, \citenamefont {Usachev}, \citenamefont
  {Zobnin}, \citenamefont {Padalka}, \citenamefont {Fortov}, \citenamefont
  {Petrov},\ and\ \citenamefont {Thomas}}]{SchwabeNJP2020}%
  \BibitemOpen
  \bibfield  {author} {\bibinfo {author} {\bibfnamefont {M.}~\bibnamefont
  {Schwabe}}, \bibinfo {author} {\bibfnamefont {S.~A.}\ \bibnamefont
  {Khrapak}}, \bibinfo {author} {\bibfnamefont {S.~K.}\ \bibnamefont
  {Zhdanov}}, \bibinfo {author} {\bibfnamefont {M.~Y.}\ \bibnamefont
  {Pustylnik}}, \bibinfo {author} {\bibfnamefont {C.}~\bibnamefont {R\"{a}th}},
  \bibinfo {author} {\bibfnamefont {M.}~\bibnamefont {Fink}}, \bibinfo {author}
  {\bibfnamefont {M.}~\bibnamefont {Kretschmer}}, \bibinfo {author}
  {\bibfnamefont {A.~M.}\ \bibnamefont {Lipaev}}, \bibinfo {author}
  {\bibfnamefont {V.~I.}\ \bibnamefont {Molotkov}}, \bibinfo {author}
  {\bibfnamefont {A.~S.}\ \bibnamefont {Schmitz}}, \bibinfo {author}
  {\bibfnamefont {M.~H.}\ \bibnamefont {Thoma}}, \bibinfo {author}
  {\bibfnamefont {A.~D.}\ \bibnamefont {Usachev}}, \bibinfo {author}
  {\bibfnamefont {A.~V.}\ \bibnamefont {Zobnin}}, \bibinfo {author}
  {\bibfnamefont {G.~I.}\ \bibnamefont {Padalka}}, \bibinfo {author}
  {\bibfnamefont {V.~E.}\ \bibnamefont {Fortov}}, \bibinfo {author}
  {\bibfnamefont {O.~F.}\ \bibnamefont {Petrov}},\ and\ \bibinfo {author}
  {\bibfnamefont {H.~M.}\ \bibnamefont {Thomas}},\ }\bibfield  {title}
  {\bibinfo {title} {Slowing of acoustic waves in electrorheological and
  string-fluid complex plasmas},\ }\href
  {https://doi.org/10.1088/1367-2630/aba91b} {\bibfield  {journal} {\bibinfo
  {journal} {New J. Phys.}\ }\textbf {\bibinfo {volume} {22}},\ \bibinfo
  {pages} {083079} (\bibinfo {year} {2020})}\BibitemShut {NoStop}%
\bibitem [{\citenamefont {D’Angelo}\ and\ \citenamefont
  {Merlino}(1996)}]{dAngeloPSS1996}%
  \BibitemOpen
  \bibfield  {author} {\bibinfo {author} {\bibfnamefont {N.}~\bibnamefont
  {D’Angelo}}\ and\ \bibinfo {author} {\bibfnamefont {R.}~\bibnamefont
  {Merlino}},\ }\bibfield  {title} {\bibinfo {title} {Current-driven
  dust-acoustic instability in a collisional plasma},\ }\href
  {https://doi.org/10.1016/s0032-0633(96)00069-4} {\bibfield  {journal}
  {\bibinfo  {journal} {Planet. Space Sci.}\ }\textbf {\bibinfo {volume}
  {44}},\ \bibinfo {pages} {1593–1598} (\bibinfo {year} {1996})}\BibitemShut
  {NoStop}%
\bibitem [{\citenamefont {Rosenberg}(1996)}]{RosenbergJVST1996}%
  \BibitemOpen
  \bibfield  {author} {\bibinfo {author} {\bibfnamefont {M.}~\bibnamefont
  {Rosenberg}},\ }\bibfield  {title} {\bibinfo {title} {Ion-dust streaming
  instability in processing plasmas},\ }\href
  {https://doi.org/10.1116/1.580157} {\bibfield  {journal} {\bibinfo  {journal}
  {J. Vac. Sci. Technol. A: Vacuum, Surfaces, and Films}\ }\textbf {\bibinfo
  {volume} {14}},\ \bibinfo {pages} {631–633} (\bibinfo {year}
  {1996})}\BibitemShut {NoStop}%
\bibitem [{\citenamefont {Ivlev}\ \emph {et~al.}(1999)\citenamefont {Ivlev},
  \citenamefont {Samsonov}, \citenamefont {Goree}, \citenamefont {Morfill},\
  and\ \citenamefont {Fortov}}]{IvlevPoP1999}%
  \BibitemOpen
  \bibfield  {author} {\bibinfo {author} {\bibfnamefont {A.~V.}\ \bibnamefont
  {Ivlev}}, \bibinfo {author} {\bibfnamefont {D.}~\bibnamefont {Samsonov}},
  \bibinfo {author} {\bibfnamefont {J.}~\bibnamefont {Goree}}, \bibinfo
  {author} {\bibfnamefont {G.}~\bibnamefont {Morfill}},\ and\ \bibinfo {author}
  {\bibfnamefont {V.~E.}\ \bibnamefont {Fortov}},\ }\bibfield  {title}
  {\bibinfo {title} {Acoustic modes in a collisional dusty plasma},\ }\href
  {https://doi.org/10.1063/1.873311} {\bibfield  {journal} {\bibinfo  {journal}
  {Phys. Plasmas}\ }\textbf {\bibinfo {volume} {6}},\ \bibinfo {pages}
  {741–750} (\bibinfo {year} {1999})}\BibitemShut {NoStop}%
\bibitem [{\citenamefont {Fortov}\ \emph {et~al.}(2000)\citenamefont {Fortov},
  \citenamefont {Khrapak}, \citenamefont {Khrapak}, \citenamefont {Molotkov},
  \citenamefont {Nefedov}, \citenamefont {Petrov},\ and\ \citenamefont
  {Torchinsky}}]{FortovPoP2000}%
  \BibitemOpen
  \bibfield  {author} {\bibinfo {author} {\bibfnamefont {V.~E.}\ \bibnamefont
  {Fortov}}, \bibinfo {author} {\bibfnamefont {A.~G.}\ \bibnamefont {Khrapak}},
  \bibinfo {author} {\bibfnamefont {S.~A.}\ \bibnamefont {Khrapak}}, \bibinfo
  {author} {\bibfnamefont {V.~I.}\ \bibnamefont {Molotkov}}, \bibinfo {author}
  {\bibfnamefont {A.~P.}\ \bibnamefont {Nefedov}}, \bibinfo {author}
  {\bibfnamefont {O.~F.}\ \bibnamefont {Petrov}},\ and\ \bibinfo {author}
  {\bibfnamefont {V.~M.}\ \bibnamefont {Torchinsky}},\ }\bibfield  {title}
  {\bibinfo {title} {Mechanism of dust-acoustic instability in a direct current
  glow discharge plasma},\ }\href {https://doi.org/10.1063/1.873954} {\bibfield
   {journal} {\bibinfo  {journal} {Phys. Plasmas}\ }\textbf {\bibinfo {volume}
  {7}},\ \bibinfo {pages} {1374–1380} (\bibinfo {year} {2000})}\BibitemShut
  {NoStop}%
\bibitem [{\citenamefont {Ratynskaia}\ \emph {et~al.}(2004)\citenamefont
  {Ratynskaia}, \citenamefont {Khrapak}, \citenamefont {Zobnin}, \citenamefont
  {Thoma}, \citenamefont {Kretschmer}, \citenamefont {Usachev}, \citenamefont
  {Yaroshenko}, \citenamefont {Quinn}, \citenamefont {Morfill}, \citenamefont
  {Petrov},\ and\ \citenamefont {Fortov}}]{RatynskaiaPRL2004}%
  \BibitemOpen
  \bibfield  {author} {\bibinfo {author} {\bibfnamefont {S.}~\bibnamefont
  {Ratynskaia}}, \bibinfo {author} {\bibfnamefont {S.}~\bibnamefont {Khrapak}},
  \bibinfo {author} {\bibfnamefont {A.}~\bibnamefont {Zobnin}}, \bibinfo
  {author} {\bibfnamefont {M.~H.}\ \bibnamefont {Thoma}}, \bibinfo {author}
  {\bibfnamefont {M.}~\bibnamefont {Kretschmer}}, \bibinfo {author}
  {\bibfnamefont {A.}~\bibnamefont {Usachev}}, \bibinfo {author} {\bibfnamefont
  {V.}~\bibnamefont {Yaroshenko}}, \bibinfo {author} {\bibfnamefont {R.~A.}\
  \bibnamefont {Quinn}}, \bibinfo {author} {\bibfnamefont {G.~E.}\ \bibnamefont
  {Morfill}}, \bibinfo {author} {\bibfnamefont {O.}~\bibnamefont {Petrov}},\
  and\ \bibinfo {author} {\bibfnamefont {V.}~\bibnamefont {Fortov}},\
  }\bibfield  {title} {\bibinfo {title} {Experimental determination of
  dust-particle charge in a discharge plasma at elevated pressures},\ }\href
  {https://doi.org/10.1103/physrevlett.93.085001} {\bibfield  {journal}
  {\bibinfo  {journal} {Phys. Rev. Lett.}\ }\textbf {\bibinfo {volume} {93}},\
  \bibinfo {pages} {085001} (\bibinfo {year} {2004})}\BibitemShut {NoStop}%
\bibitem [{\citenamefont {Piel}\ \emph {et~al.}(2008)\citenamefont {Piel},
  \citenamefont {Arp}, \citenamefont {Klindworth},\ and\ \citenamefont
  {Melzer}}]{PielPRE2008}%
  \BibitemOpen
  \bibfield  {author} {\bibinfo {author} {\bibfnamefont {A.}~\bibnamefont
  {Piel}}, \bibinfo {author} {\bibfnamefont {O.}~\bibnamefont {Arp}}, \bibinfo
  {author} {\bibfnamefont {M.}~\bibnamefont {Klindworth}},\ and\ \bibinfo
  {author} {\bibfnamefont {A.}~\bibnamefont {Melzer}},\ }\bibfield  {title}
  {\bibinfo {title} {Obliquely propagating dust-density waves},\ }\href
  {https://doi.org/10.1103/physreve.77.026407} {\bibfield  {journal} {\bibinfo
  {journal} {Phys. Rev. E}\ }\textbf {\bibinfo {volume} {77}},\ \bibinfo
  {pages} {026407} (\bibinfo {year} {2008})}\BibitemShut {NoStop}%
\bibitem [{\citenamefont {Yaroshenko}\ \emph {et~al.}(2019)\citenamefont
  {Yaroshenko}, \citenamefont {Khrapak}, \citenamefont {Pustylnik},
  \citenamefont {Thomas}, \citenamefont {Jaiswal}, \citenamefont {Lipaev},
  \citenamefont {Usachev}, \citenamefont {Petrov},\ and\ \citenamefont
  {Fortov}}]{YaroshenkoPoP2019}%
  \BibitemOpen
  \bibfield  {author} {\bibinfo {author} {\bibfnamefont {V.~V.}\ \bibnamefont
  {Yaroshenko}}, \bibinfo {author} {\bibfnamefont {S.~A.}\ \bibnamefont
  {Khrapak}}, \bibinfo {author} {\bibfnamefont {M.~Y.}\ \bibnamefont
  {Pustylnik}}, \bibinfo {author} {\bibfnamefont {H.~M.}\ \bibnamefont
  {Thomas}}, \bibinfo {author} {\bibfnamefont {S.}~\bibnamefont {Jaiswal}},
  \bibinfo {author} {\bibfnamefont {A.~M.}\ \bibnamefont {Lipaev}}, \bibinfo
  {author} {\bibfnamefont {A.~D.}\ \bibnamefont {Usachev}}, \bibinfo {author}
  {\bibfnamefont {O.~F.}\ \bibnamefont {Petrov}},\ and\ \bibinfo {author}
  {\bibfnamefont {V.~E.}\ \bibnamefont {Fortov}},\ }\bibfield  {title}
  {\bibinfo {title} {Excitation of low-frequency dust density waves in flowing
  complex plasmas},\ }\href {https://doi.org/10.1063/1.5097128} {\bibfield
  {journal} {\bibinfo  {journal} {Phys. Plasmas}\ }\textbf {\bibinfo {volume}
  {26}},\ \bibinfo {pages} {053702} (\bibinfo {year} {2019})}\BibitemShut
  {NoStop}%
\bibitem [{\citenamefont {Khrapak}\ and\ \citenamefont
  {Yaroshenko}(2020)}]{KhrapakPPCF2020}%
  \BibitemOpen
  \bibfield  {author} {\bibinfo {author} {\bibfnamefont {S.}~\bibnamefont
  {Khrapak}}\ and\ \bibinfo {author} {\bibfnamefont {V.}~\bibnamefont
  {Yaroshenko}},\ }\bibfield  {title} {\bibinfo {title} {Ion drift instability
  in a strongly coupled collisional complex plasma},\ }\href
  {https://doi.org/10.1088/1361-6587/aba7f8} {\bibfield  {journal} {\bibinfo
  {journal} {Plasma Phys. Control. Fusion}\ }\textbf {\bibinfo {volume} {62}},\
  \bibinfo {pages} {105006} (\bibinfo {year} {2020})}\BibitemShut {NoStop}%
\bibitem [{\citenamefont {Bajaj}\ \emph {et~al.}(2022)\citenamefont {Bajaj},
  \citenamefont {Khrapak}, \citenamefont {Yaroshenko},\ and\ \citenamefont
  {Schwabe}}]{BajajPRE2022}%
  \BibitemOpen
  \bibfield  {author} {\bibinfo {author} {\bibfnamefont {P.}~\bibnamefont
  {Bajaj}}, \bibinfo {author} {\bibfnamefont {S.}~\bibnamefont {Khrapak}},
  \bibinfo {author} {\bibfnamefont {V.}~\bibnamefont {Yaroshenko}},\ and\
  \bibinfo {author} {\bibfnamefont {M.}~\bibnamefont {Schwabe}},\ }\bibfield
  {title} {\bibinfo {title} {Spatial distribution of dust density wave
  properties in fluid complex plasmas},\ }\href
  {https://doi.org/10.1103/physreve.105.025202} {\bibfield  {journal} {\bibinfo
   {journal} {Phys. Rev. E}\ }\textbf {\bibinfo {volume} {105}},\ \bibinfo
  {pages} {025202} (\bibinfo {year} {2022})}\BibitemShut {NoStop}%
\bibitem [{\citenamefont {Khrapak}(2015)}]{KhrapakPPCF2015}%
  \BibitemOpen
  \bibfield  {author} {\bibinfo {author} {\bibfnamefont {S.~A.}\ \bibnamefont
  {Khrapak}},\ }\bibfield  {title} {\bibinfo {title} {Thermodynamics of
  {Y}ukawa systems and sound velocity in dusty plasmas},\ }\href
  {https://doi.org/10.1088/0741-3335/58/1/014022} {\bibfield  {journal}
  {\bibinfo  {journal} {Plasma Phys. Control. Fusion}\ }\textbf {\bibinfo
  {volume} {58}},\ \bibinfo {pages} {014022} (\bibinfo {year}
  {2015})}\BibitemShut {NoStop}%
\bibitem [{\citenamefont {Khrapak}\ and\ \citenamefont
  {Couedel}(2020)}]{KhrapakPRE09_2020}%
  \BibitemOpen
  \bibfield  {author} {\bibinfo {author} {\bibfnamefont {S.}~\bibnamefont
  {Khrapak}}\ and\ \bibinfo {author} {\bibfnamefont {L.}~\bibnamefont
  {Couedel}},\ }\bibfield  {title} {\bibinfo {title} {Dispersion relations of
  {Y}ukawa fluids at weak and moderate coupling},\ }\href
  {https://doi.org/10.1103/physreve.102.033207} {\bibfield  {journal} {\bibinfo
   {journal} {Phys. Rev. E}\ }\textbf {\bibinfo {volume} {102}},\ \bibinfo
  {pages} {033207} (\bibinfo {year} {2020})}\BibitemShut {NoStop}%
\bibitem [{\citenamefont {Silvestri}\ \emph {et~al.}(2019)\citenamefont
  {Silvestri}, \citenamefont {Kalman}, \citenamefont {Donkó}, \citenamefont
  {Hartmann}, \citenamefont {Rosenberg}, \citenamefont {Golden},\ and\
  \citenamefont {Kyrkos}}]{SilvestriPRE2019}%
  \BibitemOpen
  \bibfield  {author} {\bibinfo {author} {\bibfnamefont {L.}~\bibnamefont
  {Silvestri}}, \bibinfo {author} {\bibfnamefont {G.}~\bibnamefont {Kalman}},
  \bibinfo {author} {\bibfnamefont {Z.}~\bibnamefont {Donkó}}, \bibinfo
  {author} {\bibfnamefont {P.}~\bibnamefont {Hartmann}}, \bibinfo {author}
  {\bibfnamefont {M.}~\bibnamefont {Rosenberg}}, \bibinfo {author}
  {\bibfnamefont {K.}~\bibnamefont {Golden}},\ and\ \bibinfo {author}
  {\bibfnamefont {S.}~\bibnamefont {Kyrkos}},\ }\bibfield  {title} {\bibinfo
  {title} {Sound speed in {Y}ukawa one-component plasmas across coupling
  regimes},\ }\href {https://doi.org/10.1103/physreve.100.063206} {\bibfield
  {journal} {\bibinfo  {journal} {Phys. Rev. E}\ }\textbf {\bibinfo {volume}
  {100}},\ \bibinfo {pages} {063206} (\bibinfo {year} {2019})}\BibitemShut
  {NoStop}%
\bibitem [{\citenamefont {Khrapak}\ and\ \citenamefont
  {Thomas}(2015{\natexlab{b}})}]{KhrapakPRE02_2015}%
  \BibitemOpen
  \bibfield  {author} {\bibinfo {author} {\bibfnamefont {S.~A.}\ \bibnamefont
  {Khrapak}}\ and\ \bibinfo {author} {\bibfnamefont {H.~M.}\ \bibnamefont
  {Thomas}},\ }\bibfield  {title} {\bibinfo {title} {Practical expressions for
  the internal energy and pressure of {Y}ukawa fluids},\ }\href
  {https://doi.org/10.1103/physreve.91.023108} {\bibfield  {journal} {\bibinfo
  {journal} {Phys. Rev. E}\ }\textbf {\bibinfo {volume} {91}},\ \bibinfo
  {pages} {023108} (\bibinfo {year} {2015}{\natexlab{b}})}\BibitemShut
  {NoStop}%
\end{thebibliography}%

\end{document}